\documentclass[reprint,
superscriptaddress,
 amsmath,amssymb,
 aps,
pra,
]{revtex4-2}

\usepackage{graphicx}
\usepackage{dcolumn}
\usepackage{bm}
\usepackage{siunitx}

\usepackage{pdfpages}
\makeatletter
\AtBeginDocument{\let\LS@rot\@undefined}
\makeatother

\begin{document}
\title{Laterally Extended States of Interlayer Excitons in Reconstructed MoSe$_2$/WSe$_2$ Heterostructures}

\def\wsi{Walter Schottky Institute and Physics Department, Technical University of Munich, Am Coulombwall 4a, 85748 Garching, Germany}
\def\mcqst{Munich Center for Quantum Science and Technology (MCQST), Schellingstr. 4, 80799 Munich, Germany}
\def\berlin{Institut für Theoretische Physik, Nichtlineare Optik und
Quantenelektronik, Technische Universität Berlin, Hardenbergstr. 36, EW 7-1, 10623
Berlin, Germany}
\def\muenster{Institute of Physics, Münster University, Wilhelm-Klemm-Str. 10, 48149 Münster, Germany}

\author{Johannes Figueiredo}\affiliation{\wsi}\affiliation{\mcqst}

\author{Marten Richter}\email[Electronic address: ]{marten.richter@tu-berlin.de}\affiliation{\berlin}

\author{Mirco Troue}
\author{Jonas Kiemle}\affiliation{\wsi}\affiliation{\mcqst}

\author{Hendrik Lambers}\affiliation{\muenster}
\author{Torsten Stiehm}\affiliation{\muenster}

\author{Takashi Taniguchi}
\affiliation{International Center for Materials Nanoarchitectonics, National Institute for Materials Science, Tsukuba 305-0044, Japan}
\author{Kenji Watanabe}
\affiliation{Research Center for Functional Materials, National Institute for Materials Science, Tsukuba 305-0044, Japan}

\author{Ursula Wurstbauer}\affiliation{\muenster}

\author{Andreas Knorr}\email[Electronic address: ]{andreas.knorr@tu-berlin.de}\affiliation{\berlin}

\author{Alexander W. Holleitner}\email[Electronic address: ]{holleitner@wsi.tum.de}\affiliation{\wsi}\affiliation{\mcqst}

\date{\today}

\begin{abstract}
Heterostructures made from 2D transition-metal dichalcogenides are known as ideal platforms to explore excitonic phenomena ranging from correlated moiré excitons to degenerate interlayer exciton ensembles. So far, it is assumed that the atomic reconstruction appearing in some of the heterostructures gives rise to a dominating localization of the exciton states. We demonstrate that excitonic states in reconstructed MoSe$_2$/WSe$_2$ heterostructures can extend well beyond the moiré periodicity of the investigated heterostructures. The results are based on real-space calculations yielding a lateral potential map for interlayer excitons within the strain-relaxed heterostructures and corresponding real-space excitonic wavefunctions. We combine the theoretical results with cryogenic photoluminescence experiments, which support the computed level structure and relaxation characteristics of the interlayer excitons.
\end{abstract}
\maketitle

\section{Introduction}

\noindent Monolayers of group-VI transition-metal dichalcogenides (TMDs) have risen to prominence in solid-state research for over a decade due to their unique light-matter interactions \cite{novoselov_2d_2016, ajayan_two-dimensional_2016, manzeli_2d_2017}. The optical response of TMDs and their van der Waals heterostructures is dominated by Coulomb-bound electron-hole pairs, known as excitons, that can be tuned by dielectric engineering, doping and heterostructuring and are subjected to rich interaction as well as spin- and multivalley physics \cite{wang_colloquium_2018, he_tightly_2014, kormanyos_kp_2015, miller_tuning_2019, xiao_coupled_2012, brotons-gisbert_interlayer_2024, geim_van_2013, rivera_observation_2015, miller_long-lived_2017}. A prominent example are interlayer excitons (IXs), which particularly form across the interface of a type-II band alignment such as in MoSe$_2$/WSe$_2$ heterostructures \cite{hong_ultrafast_2014, rivera_observation_2015, wilson_determination_2017, miller_long-lived_2017, jauregui_electrical_2019}. The corresponding spatial separation of the electron and hole in adjacent layers gives rise to an increased out-of-plane electric dipole and a reduction of their wavefunction's spatial overlap. The latter significantly enhances the radiative lifetimes of the IXs to the order of tens to hundreds of nanoseconds \cite{miller_long-lived_2017, rivera_observation_2015}. Moreover, lateral moiré superlattices emerge in lattice-mismatched and/or twisted TMD hetero-bilayers and -trilayers \cite{cao_unconventional_2018, jung_ab_2014, weston_atomic_2020, rosenberger_twist_2020}. At small twist angles and/or small lattice mismatches, reconstructions can lead to the formation of lateral domains with a rather constant atomic registry and a corresponding potential landscape for electrons and holes, which typically result in strongly localized excitons \cite{zhang_interlayer_2017, jin_observation_2019, weston_atomic_2020, rosenberger_twist_2020, zhao_excitons_2023}. While the long lifetimes and the out-of-plane electric dipole position IXs as promising candidates for studying many-body phenomena across a large range of density regimes \cite{fogler_high-temperature_2014}, many findings are assumed to be limited by localization effects \cite{alexeev_resonantly_2019, jin_observation_2019,steinhoff_exciton-exciton_2024, brotons-gisbert_moire-trapped_2021, brotons-gisbert_interlayer_2024,lagoin_key_2021}.\\ 
\noindent The present work aims to explain the impact of reconstruction on the excitonic wavefunction in real space. A particular question is whether spatially extended IX states can evolve in reconstructed heterostructures at certain twist angles. Generally, the amplitude of the periodic potential landscape changes as a function of the twist angle. Recent studies suggest much lower energy amplitudes in the potential landscape for IXs in H-type MoSe$_2$/WSe$_2$ heterostructures (twist angle close to 60°) than in R-type ones (twist angle close to 0°) \cite{nielsen_accurate_2023}. We combine theoretical calculations of excitonic states in reconstructed MoSe$_2$/WSe$_2$ heterostructures close to 60° with experimental photoluminescence measurements on correspondingly designed samples. The calculations yield a two-dimensional potential landscape for the IXs within the relaxed heterostructures and a real-space representation of the IX center of mass wavefunctions of electrons and holes, as well as theoretical linear absorption and photoluminescence spectra of the IXs in the energy and time domain. The theory is formulated in real space, covering several moiré unit cells to simulate a spatially inhomogeneous exciton distribution.
Most importantly, the lateral extension of the wavefunction of the first IX eigenstates suggests delocalized excitons over more than hundreds of nanometers, particularly for twist angles larger than 1.3° with respect to 60°. Therefore, the delocalized IXs extend well beyond the moiré periodicity and are impacted by potential fluctuations within the two-dimensional plane. Moreover, the energetically lowest states located in different moiré unit cells form a single luminescence peak, which particularly dominates the calculated photoluminescence in the quasi-equilibrium regime. At smaller twist angles with respect to 60°, the calculated depth of the exciton potential increases, leading to a distribution of clearly localized exciton states, as it is consistent with earlier work \cite{seyler_signatures_2019, zhang_twist-angle_2020}. Moreover, depending on the twist angle, the type of localized states covers quantum dot-like states to networks of quantum wire-like states, again in agreement with literature \cite{brotons-gisbert_interlayer_2024}. A relatively small twist angle variation can utterly change the nature of the exciton states, forming the described resonances in photoluminescence for small twist angles close to 60°. The calculations are consistent with the presented photoluminescence experiments performed on several samples. We present low-power and time-resolved photoluminescence spectra of IXs at cryogenic temperatures, suggesting that the investigated exciton ensembles are in a dilute regime. The temporal evolution of the photoluminescence after excitation shows higher IX luminescence peaks with a distinct spectrum, which is consistent with the theoretical results. The overall observed Lorentzian-type lineshape supports the interpretation of a spatially extended IX ground state within the plane of the heterostructure for twist angles close to 60°.\\

\section{Results}

\begin{figure}
    \centering
    \includegraphics[scale=0.93]{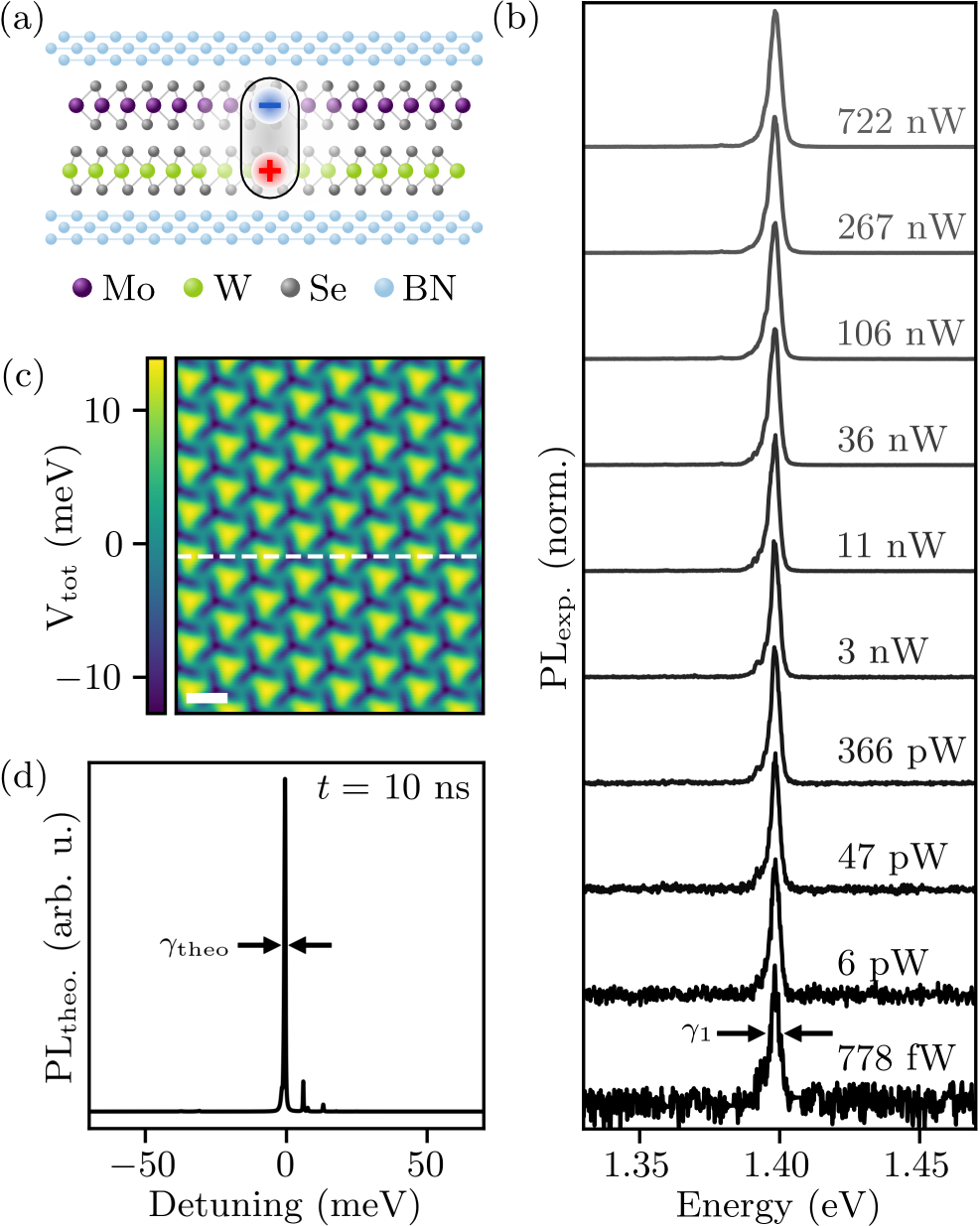}
    \caption{(a) Schematic side-view of a MoSe$_2$/WSe$_2$ heterostructure encapsulated in hBN, featuring an interlayer exciton (IX). (b) Power series of experimental photoluminescence spectra at a bath temperature of $\mathrm{T_{bath}^{exp}}=\SI{1.65}{\kelvin}$. The IXs emit light at $\mathrm{E_{IX}} = \SI{1.398}{\electronvolt}$ with a singular Lorentzian-type shape with a full-width-at-half-maximum (FWHM) $\gamma_1$. The lineshape is retained down to the lowest investigated excitation powers. (c) Simulated total interlayer potential $\mathrm{V_{tot}}$ landscape within the $x$-$y$ plane of a strain-relaxed, reconstructed bilayer at a twist angle of 1.3° wrt. 60°. The dashed line highlights a certain crystal direction, as discussed in the text. Scale bar marks \SI{10}{\micro\meter}. (d) Averaged theoretical photoluminescence spectrum with an apparent FWHM of $\gamma_{\mathrm{theo}}$ at a time delay of \SI{10}{\nano\second} after the excitation and a theoretical bath temperature of $\mathrm{T_{bath}^{theo}}=\SI{1.65}{\kelvin}$. Note that a zero detuning describes the lowest 1s IX energy without any influence from strain potentials.}
    \label{fig1}
\end{figure}

\noindent We experimentally investigate three different samples with similar photoluminescence properties, all of which consist of an hBN-encapsulated MoSe$_2$/WSe$_2$ H-type heterostructure at a twist angle close to 60°, as determined independently by magneto-photoluminescence measurements. For the rest of the manuscript, all twist angles mentioned are given as a relative deviation from the ideal 60° of a typical H-type heterostructure. Fig. \ref{fig1}(a) depicts a scheme of the investigated heterostructure highlighting an IX, where the hole and the electron reside in the different TMD monolayers. We utilize a pump laser at an energy of $\mathrm{E_{laser}} = \SI{1.94}{eV}$ to excite the charge carriers in both monolayers, thus forming IXs. Fig. \ref{fig1}(b) shows the resulting photoluminescence spectra for excitation powers from \SI{722}{\nano\watt} down to \SI{778}{\femto\watt}, where the luminescence signal is close to the overall noise floor. The single luminescence peak at $\mathrm{E_{IX}} = \SI{1.398}{\electronvolt}$ is interpreted as the recombination of IXs, and it keeps a sharp Lorentzian-type lineshape with a FWHM $\approx \SI{4}{\milli\electronvolt}$ throughout the investigated range of excitation powers.
Fig. S1 shows the power series of photoluminescence spectra on the two other samples (cf. supplementary info). For all three samples, the photon count of the low-power spectra suggests a small IX density in the dilute, single-exciton regime. Moreover, the lack of a blue shift of the emission energy with increasing pump power supports this interpretation of having IXs within the dilute regime where mutual interactions between excitons can be neglected. Given the long photoluminescence lifetime of several tens of ns, we interpret the underlying IX states as the excitonic ground states at the chosen positions of the three samples, respectively \cite{sigl_signatures_2020, troue_extended_2023}.\\
\noindent In a next step, we complement the experimental results with theoretical calculations of H-type heterostacks of MoSe$_2$ and WSe$_2$ monolayers. The underlying theoretical model including the used parameters is described in \cite{richter_theory_2024}, which contains calculations of linear spectra for R-type heterostructures. The model starts with a continuum mechanics theory to describe the lattice relaxation leading to reconstruction (based on \cite{enaldiev_stacking_2020, zhao_excitons_2023, richter_theory_2024}). The calculated displacement and strain fields form a potential for electrons and holes, which allows the calculation of interlayer states. Instead of the frequently used quasi-momentum space, the exciton states are calculated in real space to capture the effects of imperfections and disorder that disrupt translational symmetry within the plane of the heterostructures \cite{richter_theory_2024, singh_localization_2017, richter_deconvolution_2018}.
The interlayer states are used as the basis for a quantum dynamic model to describe the IXs, including exciton-phonon scattering. The quantum dynamic calculation incorporates rates using a Born-Markov approximation after polaron transformation to account for multi-phonon processes \cite{richter_theory_2024}. The calculation of optical spectra uses parameters from \cite{enaldiev_stacking_2020, khatibi_impact_2018, wu_theory_2018, tran_evidence_2019, ruiz-tijerina_theory_2020, fallahazad_shubnikov--haas_2016, mostaani_diffusion_2017, jin_intrinsic_2014, shree_observation_2018}. The utilized approach enables us to study the structure of both localized and spatially extended exciton states with respect to the center of mass motion of electrons and holes.
\begin{figure}
    \centering
    \includegraphics[scale=0.95]{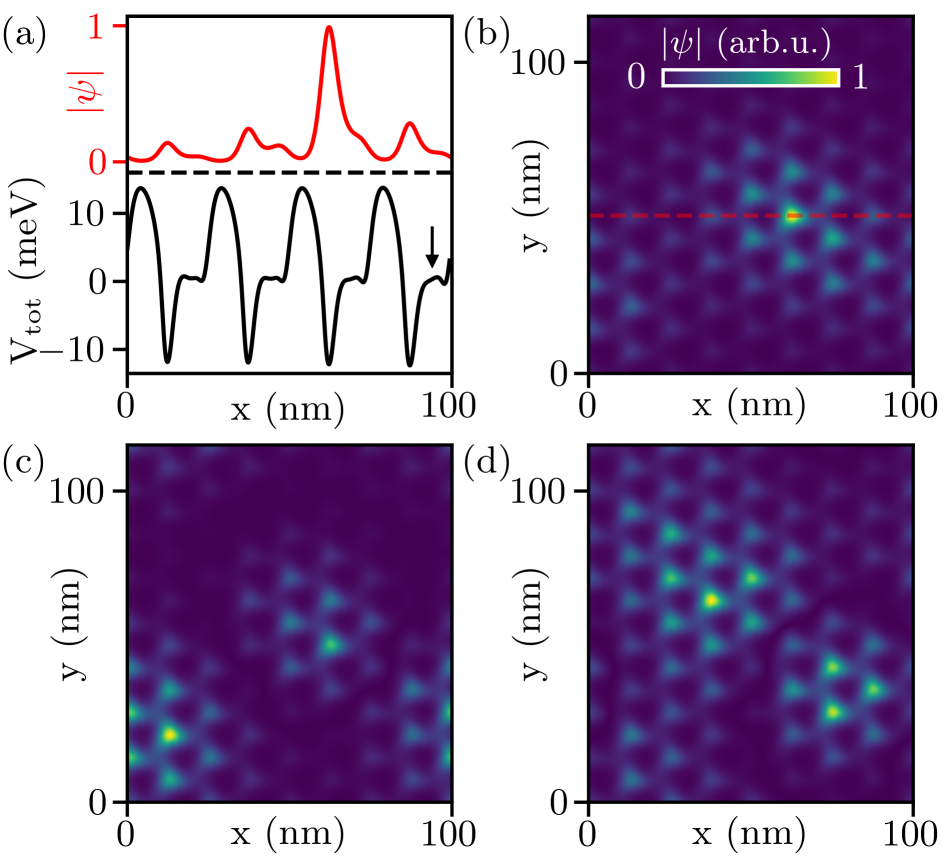}
    \caption{(a) Top curve: Absolute value of the center-of-mass (COM) wavefunction $\Psi$ for the first COM IX eigenstate along the crystallographic direction of a reconstructed bilayer at 1.3° twist angle as shown in Fig. \ref{fig1}(c). Bottom curve: Corresponding total IX potential along the same direction. The arrow indicates one of many small variations of $V_{tot}$ caused by the introduced imperfections within the simulated sample. (b-d) The absolute value of the COM wavefunction of the first three IX eigenstates in real space. The dashed red line in (b) corresponds to the $x$-axis in (a).}
    \label{fig2}
\end{figure}
\noindent The calculation starts with the displacement field of the MoSe$_2$ and WSe$_2$ monolayers forming the reconstructed lattice's top and bottom layers. The corresponding displacement field $\mathrm{\mathbf{u}_{b/t}}(\mathbf{r})$ of the bottom and the top layer is obtained by minimizing the intra- and interlayer lattice energy \cite{enaldiev_stacking_2020,zhao_excitons_2023,richter_theory_2024}, usually for an 8 x 4 supercell of the moiré structure \cite{richter_theory_2024}. We do not aim for perfect minimization but leave some residual error to simulate disorder present in the experimental samples.
\\
\\
\\
The relative interlayer displacement leads to a potential $\mathrm{V^{inter}_{e/h}}(\mathbf{r})$ for electrons and holes \cite{wu_theory_2018}. Additionally, the intralayer strain created by lattice reconstruction leads to another contribution to the potential $\mathrm{V^{intra}_{e/h}}(\mathbf{r})$ acting on electrons and holes. We combine both potentials into a total potential $\mathrm{V_{tot}}$. \\
For an exemplary twist angle deviation of 1.3°, the calculated potential landscape $\mathrm{V_{tot}}$ for IXs exhibits a periodic pattern [Fig. \ref{fig1}(c)]. The potential landscape matches experimental findings of a kagome-like lattice rearrangement dominated by hexagonal areas \cite{weston_atomic_2020, rosenberger_twist_2020}. Moreover, the calculations predict the potential to feature shallow minima when compared to R-type heterostructures of the same materials, e.g. of around \SI{-12}{\milli\electronvolt} for 1.3°. Our theory also provides photoluminescence spectra (see below for the specific calculation of the spectra), which again consider the exciton-phonon interactions in the calculation. Exemplarily, Fig. \ref{fig1}(d) shows the calculated photoluminescence ($\mathrm{PL_{theo}}$) for the already discussed 1.3° case at a time delay of \SI{10}{\nano\second} after initial injection of interlayer excitons and a theoretical temperature of $\mathrm{T_{bath}^{theo}}=\SI{1.65}{\kelvin}$. The time delay is long enough for the simulation to mimic a quasi-equilibrium near steady-state photoluminescence before a complete radiative recombination. \\
Generally, the exciton wavefunctions in real space must be calculated as a function of the spatial coordinates to understand the lateral characteristics of the excitonic states within the reconstructed heterostructures, including disorder. We start with a factorization of the exciton wavefunction $\Psi(\mathrm{\mathbf{r}_e,\mathbf{r}_h})=\psi(\mathbf{R})\phi(\mathbf{r})$ into the relative $\phi(\mathbf{r})$ and the center-of-mass (COM) $\psi(\mathbf{R})$ part. We obtain the 1s relative wavefunction $\mathrm{\phi_{1s}}(\mathbf{r})$ after solving the Schrödinger equation using finite differences and calculate the potential $\mathrm{V_{COM}}(\mathbf{R})$ acting on the COM wavefunction by convoluting the electron and hole potentials with $\mathrm{\phi_{1s}}(\mathbf{r})$ (cf. \cite{zimmermann_chapter_2003,richter_theory_2024}). We numerically compute the lowest 2500 eigenstates and eigenstates of the COM Schrödinger equation for the COM wavefunction using finite differences as the basis for the COM exciton states. If the 2500 eigenstates do not cover the spectroscopically relevant states, we choose a smaller supercell (4 x 2 instead of 8 x 4). For the 1.3° example, Fig. \ref{fig2}(a) shows both the interlayer potential and the absolute value of the wavefunction of the first IX eigenstate along the white, dashed line of the reconstructed bilayer shown in Fig. \ref{fig1}(c).
Figs. \ref{fig2}(b-d) give the absolute values of the wavefunctions $\Psi$ for the first three COM IX eigenstates as a function of the spatial coordinates $x$ and $y$. Each wavefunction varies in space because of the impact of the small spatial imperfections of $\mathrm{V_{tot}}$ within the plane of the heterostructure [cf. arrow in Fig. \ref{fig2}(a) and details in the supplementary Fig. S2]. For comparison, the red dashed line in Fig. \ref{fig2}(b) highlights the direction which resembles the $x$-axis of Fig. \ref{fig2}(a).\\
\\
\\
\\
In real experimental samples, there are often small twist angle deviations across the lateral extension of the sample. Therefore, we compare our results at 1.3° to the ones at a relative twist angle of 1.1° and 0.9°, which may manifest in other parts of a large enough sample. Figs. \ref{fig3}(a)-(c) depict the absolute value of the real space COM wavefunction $\Psi$ for each of the twist angles 1.3°, 1.1°, and 0.9°. Our results suggest that the qualitative nature of the lowest energy state changes from the 1.3° configuration towards slightly smaller twist angles. Particularly for 0.9°, the wavefunction seems to be localized within one moiré unit cell. To quantitatively capture this observation, Figs. \ref{fig3}(d)-(f) depict the lateral size (calculated as $\sqrt{\langle (\mathbf{R} - \langle \mathbf{R} \rangle )^2 \rangle}$) of the COM wavefunction for the 2500 lowest IX eigenstates for each of the three twist angles, as it is plotted as a function of the exciton energy excluding polaron shifts. Note that the maximal possible size in Fig. \ref{fig3}(d)-(f) is limited to the computational domain, so reaching the maximum means effectively a full delocalization of the corresponding state.\\
The calculated wavefunction extensions and an inspection of the COM wavefunction in real space show that the lowest energy eigenstate at 1.3° is already delocalized [Fig. \ref{fig3}(a) and red dot in Fig. \ref{fig3}(d)] and is thus not confined to one moiré unit cell (dashed line with expected moiré periodicity $\mathrm{a^{1.3^{\circ}}_M} =\SI{14.3}{\nano\metre}$) \cite{hermann_periodic_2012}. Such delocalization is not the case for smaller twist angles, as exemplarily shown in Fig. \ref{fig3}(e) and (f), where the first states are small and rather localized inside a moiré unit cell at the respective angles ($\mathrm{a^{1.1^{\circ}}_M} =\SI{16.9}{\nano\metre}$ and $\mathrm{a^{0.9^{\circ}}_M} =\SI{20.5}{\nano\metre}$). 
We note that comparing the wavefunction's size to the calculated moiré periodicity as an absolute number is a simplification, ignoring details about the shape of the wavefunction within the two-dimensional plane. For instance, some of the 2500 states for the smaller twist angles, e.g. 0.9°, exhibit rather one-dimensional quantum wire-like states along ridges of the potential landscape (cf. supplementary Fig. S3). Nonetheless, for most states and in general, the size given in Fig. \ref{fig3}(d)-(f) is a good indication of the wavefunction's (de)localization.\\
\begin{figure*}
    \centering
    \includegraphics[scale=0.95]{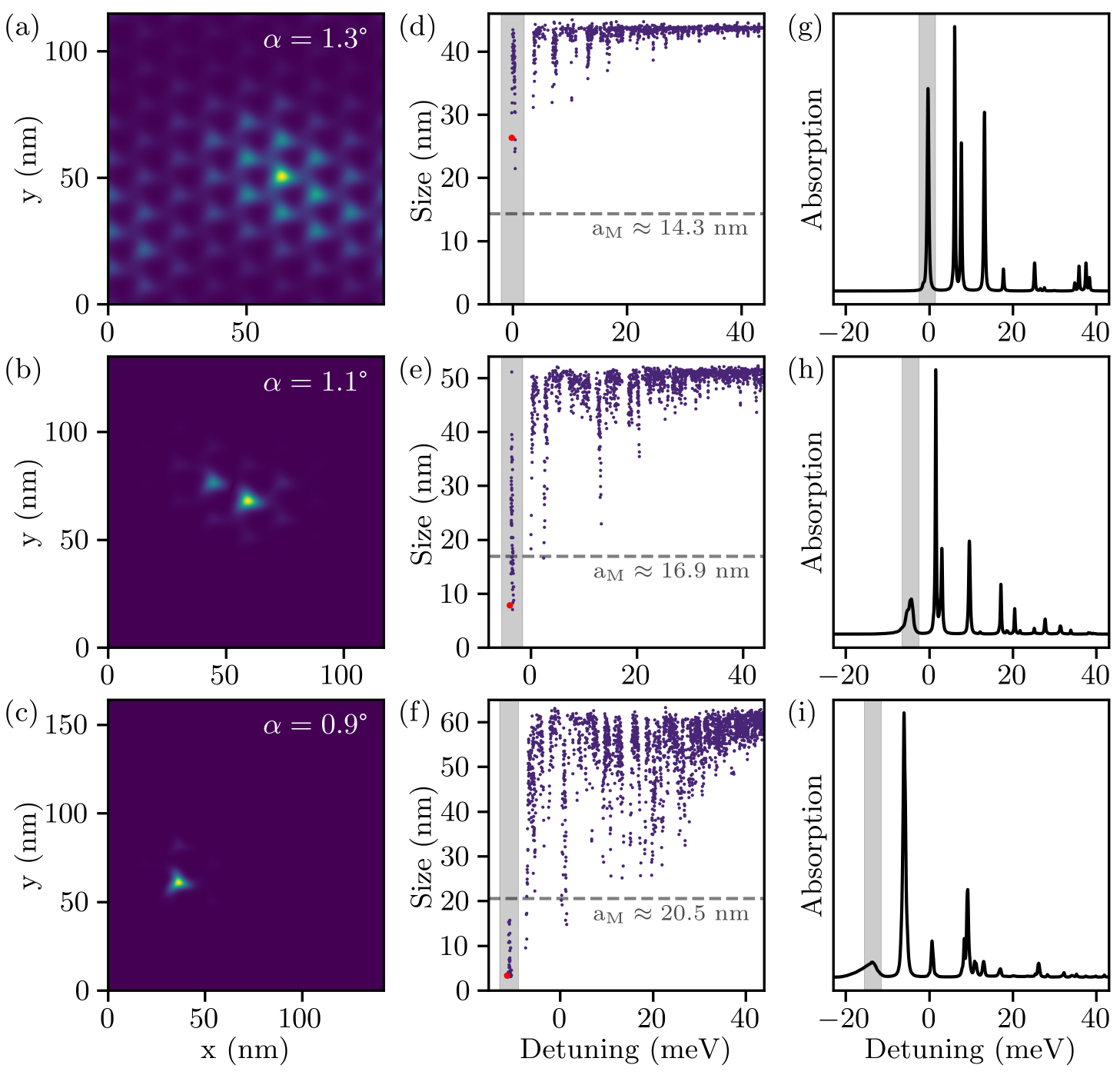}
    \caption{Spatial extension of the excitonic states and absorption spectra for a varying twist angle. (a)-(c): Absolute value of the COM wavefunction of the first IX eigenstate in real space for a relative twist angle of (a) $ \alpha =$1.3°, (b) 1.1°, and (c) 0.9°. (d) In-plane size of the wavefunction with some adjustments for the periodic computation domain. We calculate the expectation value for all 2500 exciton eigenstates vs. their energy for 1.3° with each dot in the graph representing one state. The first state, as in (a), is highlighted in red. The dashed horizontal line indicates the size of the moiré cell for the specific twist angle. (e) and (f): Corresponding representations for 1.1° and 0.9°. (g) Calculated linear absorption spectrum for the computed states at 1.3°. Gray area highlights the spectral range of the first states, as indicated by the gray area in (b). (h) and (i): Similar representations for 1.1° and 0.9°. Note that a zero detuning describes the lowest 1s IX energy without any influence from strain potentials. The shift to lower energies is consistent with a trend to exciton localization for smaller relative twist angles.}
    \label{fig3}
\end{figure*}
\noindent Next, we discuss the calculated linear absorption spectrum of the computed states. In general, the temporal Fourier transform of the dipole-dipole correlation function $\mathrm{tr}(\sigma^-_\alpha(t)\sigma^+_\alpha(0)\rho)$ (with $\sigma^+_\alpha=|\alpha\rangle\langle g|$) yields the absorption lineshape $L_\alpha(\omega-E_\alpha)$ for an exciton state $\alpha$ centered around the exciton to ground state energy $E_\alpha$. After applying the polaron transformation from the framework of \cite{richter_theory_2024}, the lineshape function becomes $\mathrm{tr}(\sigma^-_\alpha(t)B_-^\alpha(t)B_+^\alpha(0)\sigma^+_\alpha(0)\rho)$, where the operators $B_\pm^\alpha$ describe the nuclear reorganization initiated after exciton-photon interaction. A linear absorption spectrum $\alpha_{pp'}(\omega)$ for the incoming and detected polarization $p$ and $p'$ can be calculated via 
\begin{align}
    \alpha_{pp'}(\omega)=\sum_{\alpha} \mathbf{D}_{\alpha g}\cdot \mathbf{e}_p \mathbf{D}_{\alpha g}^*\cdot \mathbf{e}_{p'}^* L_\alpha(\omega-E_\alpha),
\end{align}
with the coupling element $\mathbf{D}_{\alpha g}=\int \mathrm{d} r \mathbf{D}(\mathbf{r})\psi_\alpha(\mathbf{r})$. Initially, before the excitation, there are no excitons present in linear absorption, and all bright exciton states can be accessed in the experiment. Figs. \ref{fig3}(g)-(i) show the linear absorption spectra for 1.3°, 1.1°, and 0.9° vs. the detuning energy, while a zero energy defines the lowest interlayer 1s exciton energy without the influence of the strain potentials.\\
The gray areas in Fig. \ref{fig3}(g)-(i) highlight the exciton states with the lowest energy for each twist angle with a finite absorption strength within the disordered ensemble. Comparing Fig. \ref{fig3}(g) to (h) and (i), it becomes obvious that the potential wells exhibit deeper minima for smaller twist angles. Moreover, our calculations imply an increasing spectral distinction between peaks with increasing twist angles. This is visible in the linear absorption spectrum and the energetic position of all 2500 states in the size plots. The increasing distinction is especially visible in the shape of the lowest peak, which transforms from the asymmetric shape at 0.9° to the pure Lorentzian form at 1.3°. The trend reflects the transition from a distribution of clearly localized states with an Urbach tail visible in the spectrum at (0.9°) \cite{urbach_long-wavelength_1953, piccardo_localization_2017}, to a mixture of delocalized and localized states (1.1°), and finally to only delocalized states (1.3° and above).\\
It is important to note that not all states within the linear absorption spectra are expected to be visible in photoluminescence, which explains the striking difference between the absorption spectrum for 1.3° as depicted in Fig. \ref{fig3}(g) and the corresponding photoluminescence spectrum introduced in Fig. \ref{fig1}(d). The reason for the difference is discussed in the following. The photoluminescence spectra are calculated via the correlation function
$\mathrm{tr}(\sigma^+_\alpha(t)\sigma^-_\alpha(0)\rho)$ (or after polaron transformation $\mathrm{tr}(B_+^\alpha(0)\sigma^+_\alpha(0)B_-^\alpha(t)\sigma^-_\alpha(t)\rho)$), where $\alpha$ indices the exciton state, which yields the photoluminescence lineshape function $L ^*(E_\alpha-\omega)$. The lineshape function for photoluminescence (mainly caused by electron-phonon interaction) is mirrored at the exciton energy $E_\alpha$ compared to the lineshape function in linear absorption. Note that the center of the lineshape function does not coincide with its maximum but with the zero phonon transition due to an imbalance of phonon emission and absorption processes.
While for linear absorption all optically active states contribute, the emission in photoluminescence is determined by the exciton density distribution $\rho_{\alpha\alpha}$, which approaches a quasi-equilibrium Boltzmann distribution for a long time.
Overall, the PL spectrum takes the form:
\begin{align}
    \mathrm{PL}_{p}(\omega)=\sum_{\alpha} \mathbf{D}_{\alpha g}\cdot \mathbf{e}_p \mathbf{D}_{\alpha g}^*\cdot \mathbf{e}_p^* L_\alpha^*(E_\alpha-\omega) \rho_{\alpha\alpha},
\end{align}
assuming that it is measured on time scales long enough for the coherence induced by the exciting laser to have already decayed, and that the approximate steady-state limit can be applied to describe the emission process. 
\begin{figure}
    \centering
    \includegraphics{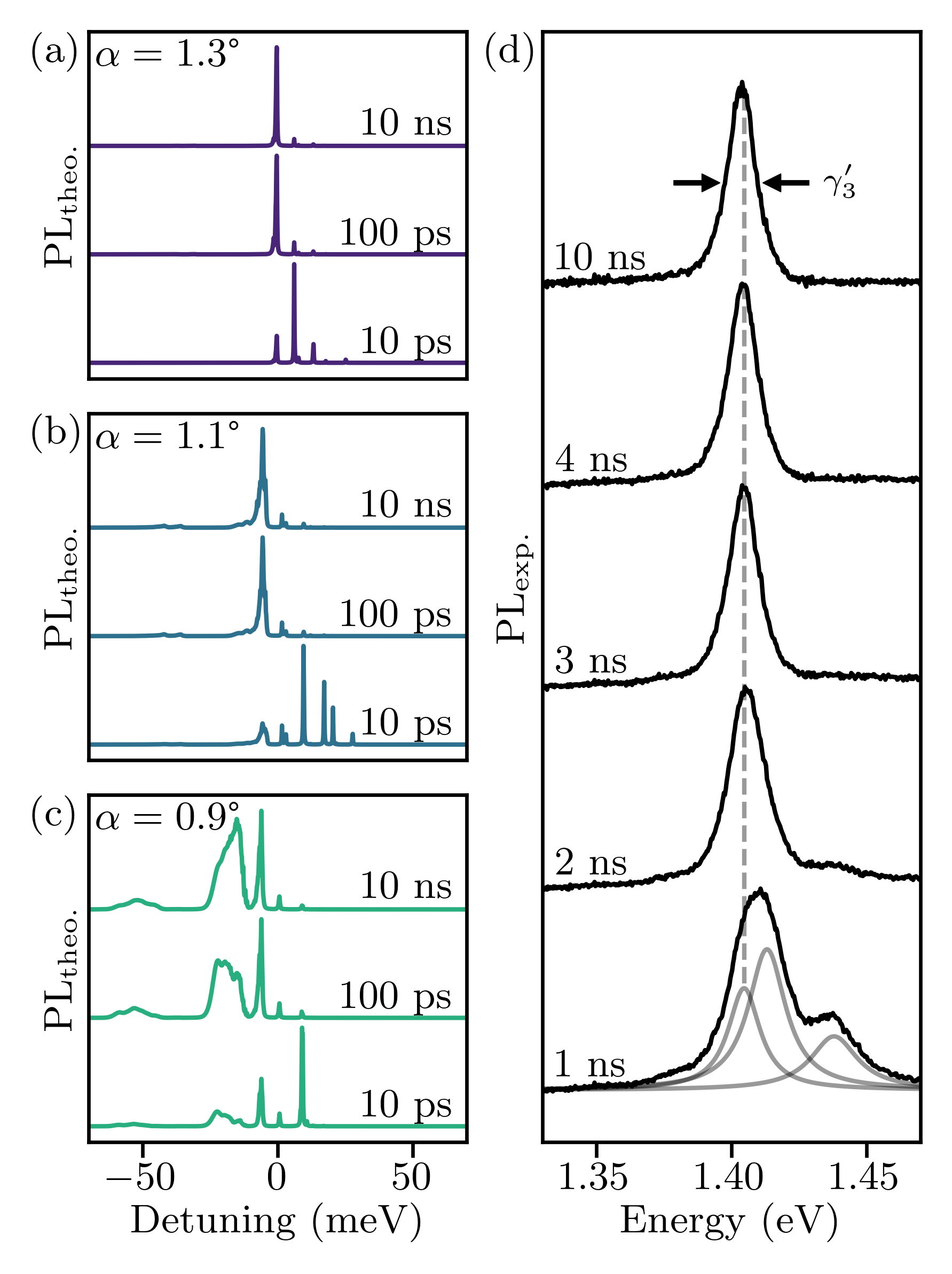}
    \caption{(a) Calculated photoluminescence spectra for a heterostructure with a relative twist angle of 1.3° at time delays ranging from 10 ps to 10 ns with respect to the excitation pulse. The temperature is $\mathrm{T_{bath}^{theo}}= \SI{100}{\milli\kelvin}$. (b) and (c): Corresponding theoretical results for 1.1° and 0.9°. (d) Experimental time-resolved photoluminescence spectra of IXs at $\mathrm{T_{bath}^{exp}}<\SI{100}{\milli\kelvin}$. All spectra are normalized.}
    \label{fig4}
\end{figure}
Fig. \ref{fig4}(a) shows calculated time-resolved photoluminescence spectra at $\mathrm{T_{bath}^{theo}}= \SI{100}{\milli\kelvin}$ for 1.3°. For the luminescence simulation, an initial Gaussian exciton density distribution of excitons at higher energies is assumed to mimic the scattering from intra-layer excitons. At short times in the \SI{10}{\pico\second} regime, the spectrum is dominated by emission lines at higher energies. These lines have a direct correspondence to the ones in the linear absorption spectra [cf. Fig. \ref{fig3}(g)]. The amplitudes vary as a function of time delay. Within a time-scale of tens to hundreds of ps, the energetically lowest state in PL increases in relative amplitude, i.e., the exciton states from the ground state peak dominate the photoluminescence at the longest times [cf. Fig. \ref{fig1}(c)]. Figures \ref{fig4}(b) and (c), show equivalent theoretical photoluminescence spectra for 1.1° and 0.9° (cf. also supplementary Fig. S4). The overall dynamics is similar for the other angles even though the nature of the involved states differs. Similar to the results for absorption, the main difference occurs in the lineshape variation and a varying spectral position of the levels. \\
In Fig. \ref{fig4}(d), we complement the theoretical work with experimental time-resolved photoluminescence on sample 3 at $\mathrm{T_{bath}^{exp}} < \SI{100}{\milli\kelvin}$. Apart from the lowest energetic emission already discussed, the first spectrum at \SI{1}{\nano\second} delay features higher-lying excitonic states, which disappear for longer delays. We fit the spectra by three Lorentzian curves with an FWHM of 14, 17, and \SI{21}{\milli\electronvolt}, respectively. Since the utilized CCD has a comparatively long gate time (2 ns), the first spectrum at a nominal time delay of 1 ns captures all events predicted by our calculations in the ps range. At later delay times of $2-\SI{10}{\nano\second}$, the lowest energetic emission dominates the photoluminescence and agrees qualitatively with the calculated spectra at 1.3°. The lines are much broader than in the measurement shown in Fig. \ref{fig1}(b). The broadening is very likely caused by a larger area contributing to the signal in this particular experimental setup (as discussed below). Consequently, the experimental signal should be described by a weighted superposition of theoretical results of different twist angles, i.e. the results complement the interpretation of the experimental data originating from areas with a reconstructed heterostructure close to 1.3° on our samples, but likely with contributions reaching down to 0.9° and lower.

\section{Discussion}

\noindent Our calculations predict a comparatively flat potential for IXs at small twist angles relative to the 60° in H-type MoSe$_2$/WSe$_2$-heterostructures [e.g. Fig. \ref{fig1}(b)] and, consequently, the emergence of delocalized IX ground states with a spatial extension of several tens of nanometers (cf. Fig. \ref{fig2}). We note that predictions of the calculated spatial extension are merely limited by the computational domain and, thus, computational power. Possible defects, samples' edges, and cracks truncate the extension of the wavefunctions, but can still be considered in the simulations \cite{richter_theory_2024}. For a particular relative twist angle of 1.3°, we observe that a spatially extended ground state dominates the calculated luminescence for long times after excitation ($>\SI{100}{\pico\second}$) [cf. Figs. \ref{fig4}(a) and \ref{fig1}(d)]. The corresponding absorption and luminescence exhibit a Lorentzian-type lineshape [cf. Figs. \ref{fig3}(g) and \ref{fig1}(d)] with a prevailing FWHM of $\mathrm{\gamma _{theo}}$ in the order of one \si{\milli\electronvolt} [Fig. \ref{fig1}(d)], fundamentally limited by the electron-phonon interaction \cite{richter_theory_2024}. The spectral width and lineshape are sensitive measures of the IXs' interaction with their environment, and thereby, as demonstrated in Fig. \ref{fig3}, also for the localization and extension of the wavefunction in a spatially slightly varying potential landscape. As localization is related to the potential depths and disorder, it will particularly influence the luminescence energy and linewidth of localized states rather than the ones of delocalized states. In other words, the spatially integrated photon absorption and emission via localized IX states reflects the potential variation of the various moiré cells, where the IXs are generated or recombine [cf. Figs. \ref{fig3}(h),(i) and \ref{fig4}(b),(c)]. Note that the simulated potential variations based on the chosen disorder parameters are in the order of 100s of \si{\micro\electronvolt} [e.g. arrow in Fig. \ref{fig2}(a) and supplementary Fig. S2] and that, in theory, a zero detuning describes the lowest 1s IX energy without any influence from strain potentials. In turn, Fig. \ref{fig3}(g) demonstrates that the Lorentzian-type lineshape evolves as soon as the impact of the potential variations is negligible, while localized states give rise to broadened non-Lorentzian lineshapes [Figs. \ref{fig3}(h),(i)]. When all previous arguments are combined, it follows that, starting at 1.3° the underlying wavefunction of the lowest energy state is delocalized in the simulations, and a deviation from a Lorentzian-type emission may reflect the impact of the potential variations in the simulated area of a heterostructure.\\
Experimentally, we observe luminescence peaks on the investigated samples with a Lorentzian-type lineshape, which prevail down to excitation powers for which the signal is only limited by the apparent noise level of the utilized optical circuitry [Fig. \ref{fig1}(b)]. Comparing $\gamma _1 = \SI{3.8}{\milli\electronvolt}$ of sample 1 to $\mathrm{\gamma _{theo}}$ [Fig. \ref{fig1}(d)], the experimental value reflects the imperfections of the investigated sample within the optical focus, but still, the observed Lorentzian-type lineshape suggests that the underlying wavefunctions are laterally extended (for $\gamma _2$, $\gamma _3$ of samples 2, 3, see supplementary info). Since we do not observe any blue-shift of the emission energy of the shown experimental luminescence [Fig. \ref{fig1}(b)], the investigated exciton ensembles can be considered to be in the dilute limit. In turn, the underlying IX states can be considered to be the lowest energy states that contribute to the luminescence in a quasi-stationary limit. Still, the time-resolved photoluminescence spectra at short time scales comprise light from states at higher energy. For instance, a higher-lying state dominates the luminescence at short time scales [Fig. \ref{fig4}(d)], which seems to be consistent with all of the shown, calculated spectra at short time scales for $\alpha$ = 1.3°, 1.1°, and 0.9° [Figs. \ref{fig4}(a)-(c)], where exciton-phonon scattering eventually leads to relaxation towards the lowest IX state. On purpose, we present calculations for more than just one relative twist angle $\alpha$ to make clear that an experiment with an optical focus of one to two micrometers collects light from presumably many specific disorder-, strain- and moiré-potential variations, as well as corresponding IX states. For instance, the rather large value of $\gamma _3 ^{\prime} = \SI{11.5}{\milli\electronvolt}$ for sample 3 as in Fig. \ref{fig4}(d) turns out to decrease to $\gamma _3 = \SI{7.5}{\milli\electronvolt}$ when the heterostructure is measured in the setup as sample 1 in Fig. \ref{fig1}(b) (cf. Fig. S1) and the lineshapes become more Lorentzian. Furthermore, if we compare the experimental lineshapes at different positions of a sample (e.g. Figs. S1, and S5) to lineshapes of the calculated photoluminescence spectra for different angles (cf. Figs. \ref{fig4}(a)-(c) and Fig. S4 at $10~\mathrm{ns}$), we tentatively conclude that the lineshape can be a qualitative indicator for assigning the twist angles under investigation.\\
On a broader perspective, we note that the mentioned electron-phonon interaction is the fundamental limit of the observed experimental FWHMs as is consistent with previous measurements on the temporal coherence of IX ensembles with a Lorentzian-type luminescence profile \cite{sigl_signatures_2020, troue_extended_2023}. The later publications report temporal coherence times of 100s of femtosecond, which translate to an FWHM in the order of a few meV as is consistent with $\gamma_{\mathrm{theo}}$. Last but not least, the deduced laterally extended wavefunctions of the IX ground states show the possibility of degenerate exciton ensembles in real samples of twisted TMD heterostructures with atomic reconstruction and small imperfections. Note, however, that the current work does not consider any exciton-exciton interactions as it would be needed to describe any transition to possible condensation and many-body phenomena of excitonic boson or fermion ensembles \cite{katzer_exciton-phonon_2023}.\\
In conclusion, our work combines theoretical calculations of excitonic states in reconstructed MoSe$_2$/WSe$_2$ bilayers close to 60° with experimental photoluminescence measurements in equivalent samples. The calculations yield a potential for IXs within the strain-relaxed bilayers and random disorder, real-space excitonic wavefunctions, linear absorption spectra, and time-resolved photoluminescence spectra, as is consistent with earlier work \cite{enaldiev_stacking_2020, zhao_excitons_2023}. Importantly, the theoretical results predict the reconstructed bilayer to exhibit a relatively flat potential landscape at 1.3° twist angle and above with an energetically lowest state featuring a single Lorentzian-type photoluminescence peak. This peak also dominates for later times in the regime of quasi-equilibrium. Both predictions are consistent with our experiments, which include low-power and time-resolved photoluminescence spectra at cryogenic temperatures. The calculated size of the wavefunction of the first COM IX eigenstates within the two-dimensional plane at 1.3° twist angle suggests a delocalization of the exciton well beyond the moiré periodicity. Such a spatial extension is in agreement with the existence of degenerate interlayer exciton ensembles in similar MoSe$_2$/WSe$_2$ bilayers, but further experiments are needed to resolve possible many-body characteristics of the IXs, such as a possible condensation and superfluidity phenomena.
\\
Acknowledgments\\
\noindent We gratefully acknowledge the German Science Foundation (DFG) for financial support via Grants HO 3324/9-2, WU 637/4-2 and 7-1, KN 427/11-2, and the clusters of excellence MCQST (EXS-2111) and e-conversion (EXS-2089), and the priority program 2244 (2DMP) via HO3324/13-2. K.W. and T.T. acknowledge support from the JSPS KAKENHI (Grant Numbers 21H05233 and 23H02052) , the CREST (JPMJCR24A5), JST and World Premier International Research Center Initiative (WPI), MEXT, Japan.
\\\\
Author contribution\\
\noindent J.F., M.T., J.K., H.L., T.S., U.W., and A.W.H. performed the experiments, M.R. and A.K. conducted the theoretical simulations, J.F., M.R., M.T., H.L., U.W., A.K., A.W.H. aligned experimental with theoretical results. J.F., J.K., T.T., K.W. provided the materials and fabricated the samples. All authors contributed to the text.
\\\\
Competing interests\\
The authors declare no competing interests.
\\\\
Data availability\\
Data is available upon reasonable request.

\newpage
\newpage

\clearpage

\bibliography{Paper_Reconstruction}

\begin{thebibliography}{49}%
\makeatletter
\providecommand \@ifxundefined [1]{%
 \@ifx{#1\undefined}
}%
\providecommand \@ifnum [1]{%
 \ifnum #1\expandafter \@firstoftwo
 \else \expandafter \@secondoftwo
 \fi
}%
\providecommand \@ifx [1]{%
 \ifx #1\expandafter \@firstoftwo
 \else \expandafter \@secondoftwo
 \fi
}%
\providecommand \natexlab [1]{#1}%
\providecommand \enquote  [1]{``#1''}%
\providecommand \bibnamefont  [1]{#1}%
\providecommand \bibfnamefont [1]{#1}%
\providecommand \citenamefont [1]{#1}%
\providecommand \href@noop [0]{\@secondoftwo}%
\providecommand \href [0]{\begingroup \@sanitize@url \@href}%
\providecommand \@href[1]{\@@startlink{#1}\@@href}%
\providecommand \@@href[1]{\endgroup#1\@@endlink}%
\providecommand \@sanitize@url [0]{\catcode `\\12\catcode `\$12\catcode `\&12\catcode `\#12\catcode `\^12\catcode `\_12\catcode `\%12\relax}%
\providecommand \@@startlink[1]{}%
\providecommand \@@endlink[0]{}%
\providecommand \url  [0]{\begingroup\@sanitize@url \@url }%
\providecommand \@url [1]{\endgroup\@href {#1}{\urlprefix }}%
\providecommand \urlprefix  [0]{URL }%
\providecommand \Eprint [0]{\href }%
\providecommand \doibase [0]{https://doi.org/}%
\providecommand \selectlanguage [0]{\@gobble}%
\providecommand \bibinfo  [0]{\@secondoftwo}%
\providecommand \bibfield  [0]{\@secondoftwo}%
\providecommand \translation [1]{[#1]}%
\providecommand \BibitemOpen [0]{}%
\providecommand \bibitemStop [0]{}%
\providecommand \bibitemNoStop [0]{.\EOS\space}%
\providecommand \EOS [0]{\spacefactor3000\relax}%
\providecommand \BibitemShut  [1]{\csname bibitem#1\endcsname}%
\let\auto@bib@innerbib\@empty
\bibitem [{\citenamefont {Novoselov}\ \emph {et~al.}(2016)\citenamefont {Novoselov}, \citenamefont {Mishchenko}, \citenamefont {Carvalho},\ and\ \citenamefont {Castro~Neto}}]{novoselov_2d_2016}%
  \BibitemOpen
  \bibfield  {author} {\bibinfo {author} {\bibfnamefont {K.~S.}\ \bibnamefont {Novoselov}}, \bibinfo {author} {\bibfnamefont {A.}~\bibnamefont {Mishchenko}}, \bibinfo {author} {\bibfnamefont {A.}~\bibnamefont {Carvalho}},\ and\ \bibinfo {author} {\bibfnamefont {A.~H.}\ \bibnamefont {Castro~Neto}},\ }\bibfield  {title} {\bibinfo {title} {{2D} materials and van der {Waals} heterostructures},\ }\href {https://doi.org/10.1126/science.aac9439} {\bibfield  {journal} {\bibinfo  {journal} {Science}\ }\textbf {\bibinfo {volume} {353}},\ \bibinfo {pages} {aac9439} (\bibinfo {year} {2016})}\BibitemShut {NoStop}%
\bibitem [{\citenamefont {Ajayan}\ \emph {et~al.}(2016)\citenamefont {Ajayan}, \citenamefont {Kim},\ and\ \citenamefont {Banerjee}}]{ajayan_two-dimensional_2016}%
  \BibitemOpen
  \bibfield  {author} {\bibinfo {author} {\bibfnamefont {P.}~\bibnamefont {Ajayan}}, \bibinfo {author} {\bibfnamefont {P.}~\bibnamefont {Kim}},\ and\ \bibinfo {author} {\bibfnamefont {K.}~\bibnamefont {Banerjee}},\ }\bibfield  {title} {\bibinfo {title} {Two-dimensional van der {Waals} materials},\ }\href {https://doi.org/10.1063/PT.3.3297} {\bibfield  {journal} {\bibinfo  {journal} {Physics Today}\ }\textbf {\bibinfo {volume} {69}},\ \bibinfo {pages} {38} (\bibinfo {year} {2016})}\BibitemShut {NoStop}%
\bibitem [{\citenamefont {Manzeli}\ \emph {et~al.}(2017)\citenamefont {Manzeli}, \citenamefont {Ovchinnikov}, \citenamefont {Pasquier}, \citenamefont {Yazyev},\ and\ \citenamefont {Kis}}]{manzeli_2d_2017}%
  \BibitemOpen
  \bibfield  {author} {\bibinfo {author} {\bibfnamefont {S.}~\bibnamefont {Manzeli}}, \bibinfo {author} {\bibfnamefont {D.}~\bibnamefont {Ovchinnikov}}, \bibinfo {author} {\bibfnamefont {D.}~\bibnamefont {Pasquier}}, \bibinfo {author} {\bibfnamefont {O.~V.}\ \bibnamefont {Yazyev}},\ and\ \bibinfo {author} {\bibfnamefont {A.}~\bibnamefont {Kis}},\ }\bibfield  {title} {\bibinfo {title} {{2D} transition metal dichalcogenides},\ }\href {https://doi.org/10.1038/natrevmats.2017.33} {\bibfield  {journal} {\bibinfo  {journal} {Nature Reviews Materials}\ }\textbf {\bibinfo {volume} {2}},\ \bibinfo {pages} {1} (\bibinfo {year} {2017})}\BibitemShut {NoStop}%
\bibitem [{\citenamefont {Wang}\ \emph {et~al.}(2018)\citenamefont {Wang}, \citenamefont {Chernikov}, \citenamefont {Glazov}, \citenamefont {Heinz}, \citenamefont {Marie}, \citenamefont {Amand},\ and\ \citenamefont {Urbaszek}}]{wang_colloquium_2018}%
  \BibitemOpen
  \bibfield  {author} {\bibinfo {author} {\bibfnamefont {G.}~\bibnamefont {Wang}}, \bibinfo {author} {\bibfnamefont {A.}~\bibnamefont {Chernikov}}, \bibinfo {author} {\bibfnamefont {M.~M.}\ \bibnamefont {Glazov}}, \bibinfo {author} {\bibfnamefont {T.~F.}\ \bibnamefont {Heinz}}, \bibinfo {author} {\bibfnamefont {X.}~\bibnamefont {Marie}}, \bibinfo {author} {\bibfnamefont {T.}~\bibnamefont {Amand}},\ and\ \bibinfo {author} {\bibfnamefont {B.}~\bibnamefont {Urbaszek}},\ }\bibfield  {title} {\bibinfo {title} {Colloquium: {Excitons} in atomically thin transition metal dichalcogenides},\ }\href {https://doi.org/10.1103/RevModPhys.90.021001} {\bibfield  {journal} {\bibinfo  {journal} {Reviews of Modern Physics}\ }\textbf {\bibinfo {volume} {90}},\ \bibinfo {pages} {021001} (\bibinfo {year} {2018})}\BibitemShut {NoStop}%
\bibitem [{\citenamefont {He}\ \emph {et~al.}(2014)\citenamefont {He}, \citenamefont {Kumar}, \citenamefont {Zhao}, \citenamefont {Wang}, \citenamefont {Mak}, \citenamefont {Zhao},\ and\ \citenamefont {Shan}}]{he_tightly_2014}%
  \BibitemOpen
  \bibfield  {author} {\bibinfo {author} {\bibfnamefont {K.}~\bibnamefont {He}}, \bibinfo {author} {\bibfnamefont {N.}~\bibnamefont {Kumar}}, \bibinfo {author} {\bibfnamefont {L.}~\bibnamefont {Zhao}}, \bibinfo {author} {\bibfnamefont {Z.}~\bibnamefont {Wang}}, \bibinfo {author} {\bibfnamefont {K.~F.}\ \bibnamefont {Mak}}, \bibinfo {author} {\bibfnamefont {H.}~\bibnamefont {Zhao}},\ and\ \bibinfo {author} {\bibfnamefont {J.}~\bibnamefont {Shan}},\ }\bibfield  {title} {\bibinfo {title} {Tightly {Bound} {Excitons} in {Monolayer} {WSe}$_2$},\ }\href {https://doi.org/10.1103/PhysRevLett.113.026803} {\bibfield  {journal} {\bibinfo  {journal} {Physical Review Letters}\ }\textbf {\bibinfo {volume} {113}},\ \bibinfo {pages} {026803} (\bibinfo {year} {2014})}\BibitemShut {NoStop}%
\bibitem [{\citenamefont {Kormányos}\ \emph {et~al.}(2015)\citenamefont {Kormányos}, \citenamefont {Burkard}, \citenamefont {Gmitra}, \citenamefont {Fabian}, \citenamefont {Zólyomi}, \citenamefont {Drummond},\ and\ \citenamefont {Fal’ko}}]{kormanyos_kp_2015}%
  \BibitemOpen
  \bibfield  {author} {\bibinfo {author} {\bibfnamefont {A.}~\bibnamefont {Kormányos}}, \bibinfo {author} {\bibfnamefont {G.}~\bibnamefont {Burkard}}, \bibinfo {author} {\bibfnamefont {M.}~\bibnamefont {Gmitra}}, \bibinfo {author} {\bibfnamefont {J.}~\bibnamefont {Fabian}}, \bibinfo {author} {\bibfnamefont {V.}~\bibnamefont {Zólyomi}}, \bibinfo {author} {\bibfnamefont {N.~D.}\ \bibnamefont {Drummond}},\ and\ \bibinfo {author} {\bibfnamefont {V.}~\bibnamefont {Fal’ko}},\ }\bibfield  {title} {\bibinfo {title} {k·p theory for two-dimensional transition metal dichalcogenide semiconductors},\ }\href {https://doi.org/10.1088/2053-1583/2/2/022001} {\bibfield  {journal} {\bibinfo  {journal} {2D Materials}\ }\textbf {\bibinfo {volume} {2}},\ \bibinfo {pages} {022001} (\bibinfo {year} {2015})}\BibitemShut {NoStop}%
\bibitem [{\citenamefont {Miller}\ \emph {et~al.}(2019)\citenamefont {Miller}, \citenamefont {Lindlau}, \citenamefont {Bommert}, \citenamefont {Neumann}, \citenamefont {Yamaguchi}, \citenamefont {Holleitner}, \citenamefont {Högele},\ and\ \citenamefont {Wurstbauer}}]{miller_tuning_2019}%
  \BibitemOpen
  \bibfield  {author} {\bibinfo {author} {\bibfnamefont {B.}~\bibnamefont {Miller}}, \bibinfo {author} {\bibfnamefont {J.}~\bibnamefont {Lindlau}}, \bibinfo {author} {\bibfnamefont {M.}~\bibnamefont {Bommert}}, \bibinfo {author} {\bibfnamefont {A.}~\bibnamefont {Neumann}}, \bibinfo {author} {\bibfnamefont {H.}~\bibnamefont {Yamaguchi}}, \bibinfo {author} {\bibfnamefont {A.}~\bibnamefont {Holleitner}}, \bibinfo {author} {\bibfnamefont {A.}~\bibnamefont {Högele}},\ and\ \bibinfo {author} {\bibfnamefont {U.}~\bibnamefont {Wurstbauer}},\ }\bibfield  {title} {\bibinfo {title} {Tuning the {Fröhlich} exciton-phonon scattering in monolayer {MoS}$_2$},\ }\href {https://doi.org/10.1038/s41467-019-08764-3} {\bibfield  {journal} {\bibinfo  {journal} {Nature Communications}\ }\textbf {\bibinfo {volume} {10}},\ \bibinfo {pages} {807} (\bibinfo {year} {2019})}\BibitemShut {NoStop}%
\bibitem [{\citenamefont {Xiao}\ \emph {et~al.}(2012)\citenamefont {Xiao}, \citenamefont {Liu}, \citenamefont {Feng}, \citenamefont {Xu},\ and\ \citenamefont {Yao}}]{xiao_coupled_2012}%
  \BibitemOpen
  \bibfield  {author} {\bibinfo {author} {\bibfnamefont {D.}~\bibnamefont {Xiao}}, \bibinfo {author} {\bibfnamefont {G.-B.}\ \bibnamefont {Liu}}, \bibinfo {author} {\bibfnamefont {W.}~\bibnamefont {Feng}}, \bibinfo {author} {\bibfnamefont {X.}~\bibnamefont {Xu}},\ and\ \bibinfo {author} {\bibfnamefont {W.}~\bibnamefont {Yao}},\ }\bibfield  {title} {\bibinfo {title} {Coupled {Spin} and {Valley} {Physics} in {Monolayers} of {MoS}$_2$ and {Other} {Group}-{VI} {Dichalcogenides}},\ }\href {https://doi.org/10.1103/PhysRevLett.108.196802} {\bibfield  {journal} {\bibinfo  {journal} {Physical Review Letters}\ }\textbf {\bibinfo {volume} {108}},\ \bibinfo {pages} {196802} (\bibinfo {year} {2012})}\BibitemShut {NoStop}%
\bibitem [{\citenamefont {Brotons-Gisbert}\ \emph {et~al.}(2024)\citenamefont {Brotons-Gisbert}, \citenamefont {Gerardot}, \citenamefont {Holleitner},\ and\ \citenamefont {Wurstbauer}}]{brotons-gisbert_interlayer_2024}%
  \BibitemOpen
  \bibfield  {author} {\bibinfo {author} {\bibfnamefont {M.}~\bibnamefont {Brotons-Gisbert}}, \bibinfo {author} {\bibfnamefont {B.~D.}\ \bibnamefont {Gerardot}}, \bibinfo {author} {\bibfnamefont {A.~W.}\ \bibnamefont {Holleitner}},\ and\ \bibinfo {author} {\bibfnamefont {U.}~\bibnamefont {Wurstbauer}},\ }\bibfield  {title} {\bibinfo {title} {Interlayer and {Moiré} excitons in atomically thin double layers: {From} individual quantum emitters to degenerate ensembles},\ }\href {https://doi.org/10.1557/s43577-024-00772-z} {\bibfield  {journal} {\bibinfo  {journal} {MRS Bulletin}\ }\textbf {\bibinfo {volume} {49}},\ \bibinfo {pages} {914} (\bibinfo {year} {2024})}\BibitemShut {NoStop}%
\bibitem [{\citenamefont {Geim}\ and\ \citenamefont {Grigorieva}(2013)}]{geim_van_2013}%
  \BibitemOpen
  \bibfield  {author} {\bibinfo {author} {\bibfnamefont {A.~K.}\ \bibnamefont {Geim}}\ and\ \bibinfo {author} {\bibfnamefont {I.~V.}\ \bibnamefont {Grigorieva}},\ }\bibfield  {title} {\bibinfo {title} {Van der {Waals} heterostructures},\ }\href {https://doi.org/10.1038/nature12385} {\bibfield  {journal} {\bibinfo  {journal} {Nature}\ }\textbf {\bibinfo {volume} {499}},\ \bibinfo {pages} {419} (\bibinfo {year} {2013})}\BibitemShut {NoStop}%
\bibitem [{\citenamefont {Rivera}\ \emph {et~al.}(2015)\citenamefont {Rivera}, \citenamefont {Schaibley}, \citenamefont {Jones}, \citenamefont {Ross}, \citenamefont {Wu}, \citenamefont {Aivazian}, \citenamefont {Klement}, \citenamefont {Seyler}, \citenamefont {Clark}, \citenamefont {Ghimire}, \citenamefont {Yan}, \citenamefont {Mandrus}, \citenamefont {Yao},\ and\ \citenamefont {Xu}}]{rivera_observation_2015}%
  \BibitemOpen
  \bibfield  {author} {\bibinfo {author} {\bibfnamefont {P.}~\bibnamefont {Rivera}}, \bibinfo {author} {\bibfnamefont {J.~R.}\ \bibnamefont {Schaibley}}, \bibinfo {author} {\bibfnamefont {A.~M.}\ \bibnamefont {Jones}}, \bibinfo {author} {\bibfnamefont {J.~S.}\ \bibnamefont {Ross}}, \bibinfo {author} {\bibfnamefont {S.}~\bibnamefont {Wu}}, \bibinfo {author} {\bibfnamefont {G.}~\bibnamefont {Aivazian}}, \bibinfo {author} {\bibfnamefont {P.}~\bibnamefont {Klement}}, \bibinfo {author} {\bibfnamefont {K.}~\bibnamefont {Seyler}}, \bibinfo {author} {\bibfnamefont {G.}~\bibnamefont {Clark}}, \bibinfo {author} {\bibfnamefont {N.~J.}\ \bibnamefont {Ghimire}}, \bibinfo {author} {\bibfnamefont {J.}~\bibnamefont {Yan}}, \bibinfo {author} {\bibfnamefont {D.~G.}\ \bibnamefont {Mandrus}}, \bibinfo {author} {\bibfnamefont {W.}~\bibnamefont {Yao}},\ and\ \bibinfo {author} {\bibfnamefont {X.}~\bibnamefont {Xu}},\ }\bibfield  {title} {\bibinfo {title} {Observation of long-lived interlayer excitons in monolayer
  {MoSe}$_2$–{WSe}$_2$ heterostructures},\ }\href {https://doi.org/10.1038/ncomms7242} {\bibfield  {journal} {\bibinfo  {journal} {Nature Communications}\ }\textbf {\bibinfo {volume} {6}},\ \bibinfo {pages} {6242} (\bibinfo {year} {2015})}\BibitemShut {NoStop}%
\bibitem [{\citenamefont {Miller}\ \emph {et~al.}(2017)\citenamefont {Miller}, \citenamefont {Steinhoff}, \citenamefont {Pano}, \citenamefont {Klein}, \citenamefont {Jahnke}, \citenamefont {Holleitner},\ and\ \citenamefont {Wurstbauer}}]{miller_long-lived_2017}%
  \BibitemOpen
  \bibfield  {author} {\bibinfo {author} {\bibfnamefont {B.}~\bibnamefont {Miller}}, \bibinfo {author} {\bibfnamefont {A.}~\bibnamefont {Steinhoff}}, \bibinfo {author} {\bibfnamefont {B.}~\bibnamefont {Pano}}, \bibinfo {author} {\bibfnamefont {J.}~\bibnamefont {Klein}}, \bibinfo {author} {\bibfnamefont {F.}~\bibnamefont {Jahnke}}, \bibinfo {author} {\bibfnamefont {A.}~\bibnamefont {Holleitner}},\ and\ \bibinfo {author} {\bibfnamefont {U.}~\bibnamefont {Wurstbauer}},\ }\bibfield  {title} {\bibinfo {title} {Long-{Lived} {Direct} and {Indirect} {Interlayer} {Excitons} in van der {Waals} {Heterostructures}},\ }\href {https://doi.org/10.1021/acs.nanolett.7b01304} {\bibfield  {journal} {\bibinfo  {journal} {Nano Letters}\ }\textbf {\bibinfo {volume} {17}},\ \bibinfo {pages} {5229} (\bibinfo {year} {2017})}\BibitemShut {NoStop}%
\bibitem [{\citenamefont {Hong}\ \emph {et~al.}(2014)\citenamefont {Hong}, \citenamefont {Kim}, \citenamefont {Shi}, \citenamefont {Zhang}, \citenamefont {Jin}, \citenamefont {Sun}, \citenamefont {Tongay}, \citenamefont {Wu}, \citenamefont {Zhang},\ and\ \citenamefont {Wang}}]{hong_ultrafast_2014}%
  \BibitemOpen
  \bibfield  {author} {\bibinfo {author} {\bibfnamefont {X.}~\bibnamefont {Hong}}, \bibinfo {author} {\bibfnamefont {J.}~\bibnamefont {Kim}}, \bibinfo {author} {\bibfnamefont {S.-F.}\ \bibnamefont {Shi}}, \bibinfo {author} {\bibfnamefont {Y.}~\bibnamefont {Zhang}}, \bibinfo {author} {\bibfnamefont {C.}~\bibnamefont {Jin}}, \bibinfo {author} {\bibfnamefont {Y.}~\bibnamefont {Sun}}, \bibinfo {author} {\bibfnamefont {S.}~\bibnamefont {Tongay}}, \bibinfo {author} {\bibfnamefont {J.}~\bibnamefont {Wu}}, \bibinfo {author} {\bibfnamefont {Y.}~\bibnamefont {Zhang}},\ and\ \bibinfo {author} {\bibfnamefont {F.}~\bibnamefont {Wang}},\ }\bibfield  {title} {\bibinfo {title} {Ultrafast charge transfer in atomically thin {MoS}$_2$/{WS}$_2$ heterostructures},\ }\href {https://doi.org/10.1038/nnano.2014.167} {\bibfield  {journal} {\bibinfo  {journal} {Nature Nanotechnology}\ }\textbf {\bibinfo {volume} {9}},\ \bibinfo {pages} {682} (\bibinfo {year} {2014})}\BibitemShut {NoStop}%
\bibitem [{\citenamefont {Wilson}\ \emph {et~al.}(2017)\citenamefont {Wilson}, \citenamefont {Nguyen}, \citenamefont {Seyler}, \citenamefont {Rivera}, \citenamefont {Marsden}, \citenamefont {Laker}, \citenamefont {Constantinescu}, \citenamefont {Kandyba}, \citenamefont {Barinov}, \citenamefont {Hine}, \citenamefont {Xu},\ and\ \citenamefont {Cobden}}]{wilson_determination_2017}%
  \BibitemOpen
  \bibfield  {author} {\bibinfo {author} {\bibfnamefont {N.~R.}\ \bibnamefont {Wilson}}, \bibinfo {author} {\bibfnamefont {P.~V.}\ \bibnamefont {Nguyen}}, \bibinfo {author} {\bibfnamefont {K.}~\bibnamefont {Seyler}}, \bibinfo {author} {\bibfnamefont {P.}~\bibnamefont {Rivera}}, \bibinfo {author} {\bibfnamefont {A.~J.}\ \bibnamefont {Marsden}}, \bibinfo {author} {\bibfnamefont {Z.~P.~L.}\ \bibnamefont {Laker}}, \bibinfo {author} {\bibfnamefont {G.~C.}\ \bibnamefont {Constantinescu}}, \bibinfo {author} {\bibfnamefont {V.}~\bibnamefont {Kandyba}}, \bibinfo {author} {\bibfnamefont {A.}~\bibnamefont {Barinov}}, \bibinfo {author} {\bibfnamefont {N.~D.~M.}\ \bibnamefont {Hine}}, \bibinfo {author} {\bibfnamefont {X.}~\bibnamefont {Xu}},\ and\ \bibinfo {author} {\bibfnamefont {D.~H.}\ \bibnamefont {Cobden}},\ }\bibfield  {title} {\bibinfo {title} {Determination of band offsets, hybridization, and exciton binding in {2D} semiconductor heterostructures},\ }\href {https://doi.org/10.1126/sciadv.1601832} {\bibfield
  {journal} {\bibinfo  {journal} {Science Advances}\ }\textbf {\bibinfo {volume} {3}},\ \bibinfo {pages} {e1601832} (\bibinfo {year} {2017})}\BibitemShut {NoStop}%
\bibitem [{\citenamefont {Jauregui}\ \emph {et~al.}(2019)\citenamefont {Jauregui}, \citenamefont {Joe}, \citenamefont {Pistunova}, \citenamefont {Wild}, \citenamefont {High}, \citenamefont {Zhou}, \citenamefont {Scuri}, \citenamefont {De~Greve}, \citenamefont {Sushko}, \citenamefont {Yu}, \citenamefont {Taniguchi}, \citenamefont {Watanabe}, \citenamefont {Needleman}, \citenamefont {Lukin}, \citenamefont {Park},\ and\ \citenamefont {Kim}}]{jauregui_electrical_2019}%
  \BibitemOpen
  \bibfield  {author} {\bibinfo {author} {\bibfnamefont {L.~A.}\ \bibnamefont {Jauregui}}, \bibinfo {author} {\bibfnamefont {A.~Y.}\ \bibnamefont {Joe}}, \bibinfo {author} {\bibfnamefont {K.}~\bibnamefont {Pistunova}}, \bibinfo {author} {\bibfnamefont {D.~S.}\ \bibnamefont {Wild}}, \bibinfo {author} {\bibfnamefont {A.~A.}\ \bibnamefont {High}}, \bibinfo {author} {\bibfnamefont {Y.}~\bibnamefont {Zhou}}, \bibinfo {author} {\bibfnamefont {G.}~\bibnamefont {Scuri}}, \bibinfo {author} {\bibfnamefont {K.}~\bibnamefont {De~Greve}}, \bibinfo {author} {\bibfnamefont {A.}~\bibnamefont {Sushko}}, \bibinfo {author} {\bibfnamefont {C.-H.}\ \bibnamefont {Yu}}, \bibinfo {author} {\bibfnamefont {T.}~\bibnamefont {Taniguchi}}, \bibinfo {author} {\bibfnamefont {K.}~\bibnamefont {Watanabe}}, \bibinfo {author} {\bibfnamefont {D.~J.}\ \bibnamefont {Needleman}}, \bibinfo {author} {\bibfnamefont {M.~D.}\ \bibnamefont {Lukin}}, \bibinfo {author} {\bibfnamefont {H.}~\bibnamefont {Park}},\ and\ \bibinfo {author} {\bibfnamefont
  {P.}~\bibnamefont {Kim}},\ }\bibfield  {title} {\bibinfo {title} {Electrical control of interlayer exciton dynamics in atomically thin heterostructures},\ }\href {https://doi.org/10.1126/science.aaw4194} {\bibfield  {journal} {\bibinfo  {journal} {Science}\ }\textbf {\bibinfo {volume} {366}},\ \bibinfo {pages} {870} (\bibinfo {year} {2019})}\BibitemShut {NoStop}%
\bibitem [{\citenamefont {Cao}\ \emph {et~al.}(2018)\citenamefont {Cao}, \citenamefont {Fatemi}, \citenamefont {Fang}, \citenamefont {Watanabe}, \citenamefont {Taniguchi}, \citenamefont {Kaxiras},\ and\ \citenamefont {Jarillo-Herrero}}]{cao_unconventional_2018}%
  \BibitemOpen
  \bibfield  {author} {\bibinfo {author} {\bibfnamefont {Y.}~\bibnamefont {Cao}}, \bibinfo {author} {\bibfnamefont {V.}~\bibnamefont {Fatemi}}, \bibinfo {author} {\bibfnamefont {S.}~\bibnamefont {Fang}}, \bibinfo {author} {\bibfnamefont {K.}~\bibnamefont {Watanabe}}, \bibinfo {author} {\bibfnamefont {T.}~\bibnamefont {Taniguchi}}, \bibinfo {author} {\bibfnamefont {E.}~\bibnamefont {Kaxiras}},\ and\ \bibinfo {author} {\bibfnamefont {P.}~\bibnamefont {Jarillo-Herrero}},\ }\bibfield  {title} {\bibinfo {title} {Unconventional superconductivity in magic-angle graphene superlattices},\ }\href {https://doi.org/10.1038/nature26160} {\bibfield  {journal} {\bibinfo  {journal} {Nature}\ }\textbf {\bibinfo {volume} {556}},\ \bibinfo {pages} {43} (\bibinfo {year} {2018})}\BibitemShut {NoStop}%
\bibitem [{\citenamefont {Jung}\ \emph {et~al.}(2014)\citenamefont {Jung}, \citenamefont {Raoux}, \citenamefont {Qiao},\ and\ \citenamefont {MacDonald}}]{jung_ab_2014}%
  \BibitemOpen
  \bibfield  {author} {\bibinfo {author} {\bibfnamefont {J.}~\bibnamefont {Jung}}, \bibinfo {author} {\bibfnamefont {A.}~\bibnamefont {Raoux}}, \bibinfo {author} {\bibfnamefont {Z.}~\bibnamefont {Qiao}},\ and\ \bibinfo {author} {\bibfnamefont {A.~H.}\ \bibnamefont {MacDonald}},\ }\bibfield  {title} {\bibinfo {title} {Ab initio theory of moiré superlattice bands in layered two-dimensional materials},\ }\href {https://doi.org/10.1103/PhysRevB.89.205414} {\bibfield  {journal} {\bibinfo  {journal} {Physical Review B}\ }\textbf {\bibinfo {volume} {89}},\ \bibinfo {pages} {205414} (\bibinfo {year} {2014})}\BibitemShut {NoStop}%
\bibitem [{\citenamefont {Weston}\ \emph {et~al.}(2020)\citenamefont {Weston}, \citenamefont {Zou}, \citenamefont {Enaldiev}, \citenamefont {Summerfield}, \citenamefont {Clark}, \citenamefont {Zólyomi}, \citenamefont {Graham}, \citenamefont {Yelgel}, \citenamefont {Magorrian}, \citenamefont {Zhou}, \citenamefont {Zultak}, \citenamefont {Hopkinson}, \citenamefont {Barinov}, \citenamefont {Bointon}, \citenamefont {Kretinin}, \citenamefont {Wilson}, \citenamefont {Beton}, \citenamefont {Fal’ko}, \citenamefont {Haigh},\ and\ \citenamefont {Gorbachev}}]{weston_atomic_2020}%
  \BibitemOpen
  \bibfield  {author} {\bibinfo {author} {\bibfnamefont {A.}~\bibnamefont {Weston}}, \bibinfo {author} {\bibfnamefont {Y.}~\bibnamefont {Zou}}, \bibinfo {author} {\bibfnamefont {V.}~\bibnamefont {Enaldiev}}, \bibinfo {author} {\bibfnamefont {A.}~\bibnamefont {Summerfield}}, \bibinfo {author} {\bibfnamefont {N.}~\bibnamefont {Clark}}, \bibinfo {author} {\bibfnamefont {V.}~\bibnamefont {Zólyomi}}, \bibinfo {author} {\bibfnamefont {A.}~\bibnamefont {Graham}}, \bibinfo {author} {\bibfnamefont {C.}~\bibnamefont {Yelgel}}, \bibinfo {author} {\bibfnamefont {S.}~\bibnamefont {Magorrian}}, \bibinfo {author} {\bibfnamefont {M.}~\bibnamefont {Zhou}}, \bibinfo {author} {\bibfnamefont {J.}~\bibnamefont {Zultak}}, \bibinfo {author} {\bibfnamefont {D.}~\bibnamefont {Hopkinson}}, \bibinfo {author} {\bibfnamefont {A.}~\bibnamefont {Barinov}}, \bibinfo {author} {\bibfnamefont {T.~H.}\ \bibnamefont {Bointon}}, \bibinfo {author} {\bibfnamefont {A.}~\bibnamefont {Kretinin}}, \bibinfo {author} {\bibfnamefont {N.~R.}\ \bibnamefont
  {Wilson}}, \bibinfo {author} {\bibfnamefont {P.~H.}\ \bibnamefont {Beton}}, \bibinfo {author} {\bibfnamefont {V.~I.}\ \bibnamefont {Fal’ko}}, \bibinfo {author} {\bibfnamefont {S.~J.}\ \bibnamefont {Haigh}},\ and\ \bibinfo {author} {\bibfnamefont {R.}~\bibnamefont {Gorbachev}},\ }\bibfield  {title} {\bibinfo {title} {Atomic reconstruction in twisted bilayers of transition metal dichalcogenides},\ }\href {https://doi.org/10.1038/s41565-020-0682-9} {\bibfield  {journal} {\bibinfo  {journal} {Nature Nanotechnology}\ }\textbf {\bibinfo {volume} {15}},\ \bibinfo {pages} {592} (\bibinfo {year} {2020})}\BibitemShut {NoStop}%
\bibitem [{\citenamefont {Rosenberger}\ \emph {et~al.}(2020)\citenamefont {Rosenberger}, \citenamefont {Chuang}, \citenamefont {Phillips}, \citenamefont {Oleshko}, \citenamefont {McCreary}, \citenamefont {Sivaram}, \citenamefont {Hellberg},\ and\ \citenamefont {Jonker}}]{rosenberger_twist_2020}%
  \BibitemOpen
  \bibfield  {author} {\bibinfo {author} {\bibfnamefont {M.~R.}\ \bibnamefont {Rosenberger}}, \bibinfo {author} {\bibfnamefont {H.-J.}\ \bibnamefont {Chuang}}, \bibinfo {author} {\bibfnamefont {M.}~\bibnamefont {Phillips}}, \bibinfo {author} {\bibfnamefont {V.~P.}\ \bibnamefont {Oleshko}}, \bibinfo {author} {\bibfnamefont {K.~M.}\ \bibnamefont {McCreary}}, \bibinfo {author} {\bibfnamefont {S.~V.}\ \bibnamefont {Sivaram}}, \bibinfo {author} {\bibfnamefont {C.~S.}\ \bibnamefont {Hellberg}},\ and\ \bibinfo {author} {\bibfnamefont {B.~T.}\ \bibnamefont {Jonker}},\ }\bibfield  {title} {\bibinfo {title} {Twist {Angle}-{Dependent} {Atomic} {Reconstruction} and {Moiré} {Patterns} in {Transition} {Metal} {Dichalcogenide} {Heterostructures}},\ }\href {https://doi.org/10.1021/acsnano.0c00088} {\bibfield  {journal} {\bibinfo  {journal} {ACS Nano}\ }\textbf {\bibinfo {volume} {14}},\ \bibinfo {pages} {4550} (\bibinfo {year} {2020})}\BibitemShut {NoStop}%
\bibitem [{\citenamefont {Zhang}\ \emph {et~al.}(2017)\citenamefont {Zhang}, \citenamefont {Chuu}, \citenamefont {Ren}, \citenamefont {Li}, \citenamefont {Li}, \citenamefont {Jin}, \citenamefont {Chou},\ and\ \citenamefont {Shih}}]{zhang_interlayer_2017}%
  \BibitemOpen
  \bibfield  {author} {\bibinfo {author} {\bibfnamefont {C.}~\bibnamefont {Zhang}}, \bibinfo {author} {\bibfnamefont {C.-P.}\ \bibnamefont {Chuu}}, \bibinfo {author} {\bibfnamefont {X.}~\bibnamefont {Ren}}, \bibinfo {author} {\bibfnamefont {M.-Y.}\ \bibnamefont {Li}}, \bibinfo {author} {\bibfnamefont {L.-J.}\ \bibnamefont {Li}}, \bibinfo {author} {\bibfnamefont {C.}~\bibnamefont {Jin}}, \bibinfo {author} {\bibfnamefont {M.-Y.}\ \bibnamefont {Chou}},\ and\ \bibinfo {author} {\bibfnamefont {C.-K.}\ \bibnamefont {Shih}},\ }\bibfield  {title} {\bibinfo {title} {Interlayer couplings, {Moiré} patterns, and {2D} electronic superlattices in {MoS}$_2$/{WSe}$_2$ hetero-bilayers},\ }\href {https://doi.org/10.1126/sciadv.1601459} {\bibfield  {journal} {\bibinfo  {journal} {Science Advances}\ }\textbf {\bibinfo {volume} {3}},\ \bibinfo {pages} {e1601459} (\bibinfo {year} {2017})}\BibitemShut {NoStop}%
\bibitem [{\citenamefont {Jin}\ \emph {et~al.}(2019)\citenamefont {Jin}, \citenamefont {Regan}, \citenamefont {Yan}, \citenamefont {Iqbal Bakti~Utama}, \citenamefont {Wang}, \citenamefont {Zhao}, \citenamefont {Qin}, \citenamefont {Yang}, \citenamefont {Zheng}, \citenamefont {Shi}, \citenamefont {Watanabe}, \citenamefont {Taniguchi}, \citenamefont {Tongay}, \citenamefont {Zettl},\ and\ \citenamefont {Wang}}]{jin_observation_2019}%
  \BibitemOpen
  \bibfield  {author} {\bibinfo {author} {\bibfnamefont {C.}~\bibnamefont {Jin}}, \bibinfo {author} {\bibfnamefont {E.~C.}\ \bibnamefont {Regan}}, \bibinfo {author} {\bibfnamefont {A.}~\bibnamefont {Yan}}, \bibinfo {author} {\bibfnamefont {M.}~\bibnamefont {Iqbal Bakti~Utama}}, \bibinfo {author} {\bibfnamefont {D.}~\bibnamefont {Wang}}, \bibinfo {author} {\bibfnamefont {S.}~\bibnamefont {Zhao}}, \bibinfo {author} {\bibfnamefont {Y.}~\bibnamefont {Qin}}, \bibinfo {author} {\bibfnamefont {S.}~\bibnamefont {Yang}}, \bibinfo {author} {\bibfnamefont {Z.}~\bibnamefont {Zheng}}, \bibinfo {author} {\bibfnamefont {S.}~\bibnamefont {Shi}}, \bibinfo {author} {\bibfnamefont {K.}~\bibnamefont {Watanabe}}, \bibinfo {author} {\bibfnamefont {T.}~\bibnamefont {Taniguchi}}, \bibinfo {author} {\bibfnamefont {S.}~\bibnamefont {Tongay}}, \bibinfo {author} {\bibfnamefont {A.}~\bibnamefont {Zettl}},\ and\ \bibinfo {author} {\bibfnamefont {F.}~\bibnamefont {Wang}},\ }\bibfield  {title} {\bibinfo {title} {Observation of moiré
  excitons in {WSe}$_2$/{WS}$_2$ heterostructure superlattices},\ }\href {https://doi.org/10.1038/s41586-019-0976-y} {\bibfield  {journal} {\bibinfo  {journal} {Nature}\ }\textbf {\bibinfo {volume} {567}},\ \bibinfo {pages} {76} (\bibinfo {year} {2019})}\BibitemShut {NoStop}%
\bibitem [{\citenamefont {Zhao}\ \emph {et~al.}(2023)\citenamefont {Zhao}, \citenamefont {Li}, \citenamefont {Huang}, \citenamefont {Rupp}, \citenamefont {Göser}, \citenamefont {Vovk}, \citenamefont {Kruchinin}, \citenamefont {Watanabe}, \citenamefont {Taniguchi}, \citenamefont {Bilgin}, \citenamefont {Baimuratov},\ and\ \citenamefont {Högele}}]{zhao_excitons_2023}%
  \BibitemOpen
  \bibfield  {author} {\bibinfo {author} {\bibfnamefont {S.}~\bibnamefont {Zhao}}, \bibinfo {author} {\bibfnamefont {Z.}~\bibnamefont {Li}}, \bibinfo {author} {\bibfnamefont {X.}~\bibnamefont {Huang}}, \bibinfo {author} {\bibfnamefont {A.}~\bibnamefont {Rupp}}, \bibinfo {author} {\bibfnamefont {J.}~\bibnamefont {Göser}}, \bibinfo {author} {\bibfnamefont {I.~A.}\ \bibnamefont {Vovk}}, \bibinfo {author} {\bibfnamefont {S.~Y.}\ \bibnamefont {Kruchinin}}, \bibinfo {author} {\bibfnamefont {K.}~\bibnamefont {Watanabe}}, \bibinfo {author} {\bibfnamefont {T.}~\bibnamefont {Taniguchi}}, \bibinfo {author} {\bibfnamefont {I.}~\bibnamefont {Bilgin}}, \bibinfo {author} {\bibfnamefont {A.~S.}\ \bibnamefont {Baimuratov}},\ and\ \bibinfo {author} {\bibfnamefont {A.}~\bibnamefont {Högele}},\ }\bibfield  {title} {\bibinfo {title} {Excitons in mesoscopically reconstructed moiré heterostructures},\ }\href {https://doi.org/10.1038/s41565-023-01356-9} {\bibfield  {journal} {\bibinfo  {journal} {Nature Nanotechnology}\ }\textbf
  {\bibinfo {volume} {18}},\ \bibinfo {pages} {572} (\bibinfo {year} {2023})}\BibitemShut {NoStop}%
\bibitem [{\citenamefont {Fogler}\ \emph {et~al.}(2014)\citenamefont {Fogler}, \citenamefont {Butov},\ and\ \citenamefont {Novoselov}}]{fogler_high-temperature_2014}%
  \BibitemOpen
  \bibfield  {author} {\bibinfo {author} {\bibfnamefont {M.~M.}\ \bibnamefont {Fogler}}, \bibinfo {author} {\bibfnamefont {L.~V.}\ \bibnamefont {Butov}},\ and\ \bibinfo {author} {\bibfnamefont {K.~S.}\ \bibnamefont {Novoselov}},\ }\bibfield  {title} {\bibinfo {title} {High-temperature superfluidity with indirect excitons in van der {Waals} heterostructures},\ }\href {https://doi.org/10.1038/ncomms5555} {\bibfield  {journal} {\bibinfo  {journal} {Nature Communications}\ }\textbf {\bibinfo {volume} {5}},\ \bibinfo {pages} {4555} (\bibinfo {year} {2014})}\BibitemShut {NoStop}%
\bibitem [{\citenamefont {Alexeev}\ \emph {et~al.}(2019)\citenamefont {Alexeev}, \citenamefont {Ruiz-Tijerina}, \citenamefont {Danovich}, \citenamefont {Hamer}, \citenamefont {Terry}, \citenamefont {Nayak}, \citenamefont {Ahn}, \citenamefont {Pak}, \citenamefont {Lee}, \citenamefont {Sohn}, \citenamefont {Molas}, \citenamefont {Koperski}, \citenamefont {Watanabe}, \citenamefont {Taniguchi}, \citenamefont {Novoselov}, \citenamefont {Gorbachev}, \citenamefont {Shin}, \citenamefont {Fal’ko},\ and\ \citenamefont {Tartakovskii}}]{alexeev_resonantly_2019}%
  \BibitemOpen
  \bibfield  {author} {\bibinfo {author} {\bibfnamefont {E.~M.}\ \bibnamefont {Alexeev}}, \bibinfo {author} {\bibfnamefont {D.~A.}\ \bibnamefont {Ruiz-Tijerina}}, \bibinfo {author} {\bibfnamefont {M.}~\bibnamefont {Danovich}}, \bibinfo {author} {\bibfnamefont {M.~J.}\ \bibnamefont {Hamer}}, \bibinfo {author} {\bibfnamefont {D.~J.}\ \bibnamefont {Terry}}, \bibinfo {author} {\bibfnamefont {P.~K.}\ \bibnamefont {Nayak}}, \bibinfo {author} {\bibfnamefont {S.}~\bibnamefont {Ahn}}, \bibinfo {author} {\bibfnamefont {S.}~\bibnamefont {Pak}}, \bibinfo {author} {\bibfnamefont {J.}~\bibnamefont {Lee}}, \bibinfo {author} {\bibfnamefont {J.~I.}\ \bibnamefont {Sohn}}, \bibinfo {author} {\bibfnamefont {M.~R.}\ \bibnamefont {Molas}}, \bibinfo {author} {\bibfnamefont {M.}~\bibnamefont {Koperski}}, \bibinfo {author} {\bibfnamefont {K.}~\bibnamefont {Watanabe}}, \bibinfo {author} {\bibfnamefont {T.}~\bibnamefont {Taniguchi}}, \bibinfo {author} {\bibfnamefont {K.~S.}\ \bibnamefont {Novoselov}}, \bibinfo {author} {\bibfnamefont
  {R.~V.}\ \bibnamefont {Gorbachev}}, \bibinfo {author} {\bibfnamefont {H.~S.}\ \bibnamefont {Shin}}, \bibinfo {author} {\bibfnamefont {V.~I.}\ \bibnamefont {Fal’ko}},\ and\ \bibinfo {author} {\bibfnamefont {A.~I.}\ \bibnamefont {Tartakovskii}},\ }\bibfield  {title} {\bibinfo {title} {Resonantly hybridized excitons in moiré superlattices in van der {Waals} heterostructures},\ }\href {https://doi.org/10.1038/s41586-019-0986-9} {\bibfield  {journal} {\bibinfo  {journal} {Nature}\ }\textbf {\bibinfo {volume} {567}},\ \bibinfo {pages} {81} (\bibinfo {year} {2019})}\BibitemShut {NoStop}%
\bibitem [{\citenamefont {Steinhoff}\ \emph {et~al.}(2024)\citenamefont {Steinhoff}, \citenamefont {Wietek}, \citenamefont {Florian}, \citenamefont {Schulz}, \citenamefont {Taniguchi}, \citenamefont {Watanabe}, \citenamefont {Zhao}, \citenamefont {Högele}, \citenamefont {Jahnke},\ and\ \citenamefont {Chernikov}}]{steinhoff_exciton-exciton_2024}%
  \BibitemOpen
  \bibfield  {author} {\bibinfo {author} {\bibfnamefont {A.}~\bibnamefont {Steinhoff}}, \bibinfo {author} {\bibfnamefont {E.}~\bibnamefont {Wietek}}, \bibinfo {author} {\bibfnamefont {M.}~\bibnamefont {Florian}}, \bibinfo {author} {\bibfnamefont {T.}~\bibnamefont {Schulz}}, \bibinfo {author} {\bibfnamefont {T.}~\bibnamefont {Taniguchi}}, \bibinfo {author} {\bibfnamefont {K.}~\bibnamefont {Watanabe}}, \bibinfo {author} {\bibfnamefont {S.}~\bibnamefont {Zhao}}, \bibinfo {author} {\bibfnamefont {A.}~\bibnamefont {Högele}}, \bibinfo {author} {\bibfnamefont {F.}~\bibnamefont {Jahnke}},\ and\ \bibinfo {author} {\bibfnamefont {A.}~\bibnamefont {Chernikov}},\ }\bibfield  {title} {\bibinfo {title} {Exciton-{Exciton} {Interactions} in {Van} der {Waals} {Heterobilayers}},\ }\href {https://doi.org/10.1103/PhysRevX.14.031025} {\bibfield  {journal} {\bibinfo  {journal} {Physical Review X}\ }\textbf {\bibinfo {volume} {14}},\ \bibinfo {pages} {031025} (\bibinfo {year} {2024})}\BibitemShut {NoStop}%
\bibitem [{\citenamefont {Brotons-Gisbert}\ \emph {et~al.}(2021)\citenamefont {Brotons-Gisbert}, \citenamefont {Baek}, \citenamefont {Campbell}, \citenamefont {Watanabe}, \citenamefont {Taniguchi},\ and\ \citenamefont {Gerardot}}]{brotons-gisbert_moire-trapped_2021}%
  \BibitemOpen
  \bibfield  {author} {\bibinfo {author} {\bibfnamefont {M.}~\bibnamefont {Brotons-Gisbert}}, \bibinfo {author} {\bibfnamefont {H.}~\bibnamefont {Baek}}, \bibinfo {author} {\bibfnamefont {A.}~\bibnamefont {Campbell}}, \bibinfo {author} {\bibfnamefont {K.}~\bibnamefont {Watanabe}}, \bibinfo {author} {\bibfnamefont {T.}~\bibnamefont {Taniguchi}},\ and\ \bibinfo {author} {\bibfnamefont {B.~D.}\ \bibnamefont {Gerardot}},\ }\bibfield  {title} {\bibinfo {title} {Moiré-{Trapped} {Interlayer} {Trions} in a {Charge}-{Tunable} {WSe}$_2$/{MoSe}$_2$ {Heterobilayer}},\ }\href {https://doi.org/10.1103/PhysRevX.11.031033} {\bibfield  {journal} {\bibinfo  {journal} {Physical Review X}\ }\textbf {\bibinfo {volume} {11}},\ \bibinfo {pages} {031033} (\bibinfo {year} {2021})}\BibitemShut {NoStop}%
\bibitem [{\citenamefont {Lagoin}\ and\ \citenamefont {Dubin}(2021)}]{lagoin_key_2021}%
  \BibitemOpen
  \bibfield  {author} {\bibinfo {author} {\bibfnamefont {C.}~\bibnamefont {Lagoin}}\ and\ \bibinfo {author} {\bibfnamefont {F.}~\bibnamefont {Dubin}},\ }\bibfield  {title} {\bibinfo {title} {Key role of the moiré potential for the quasicondensation of interlayer excitons in van der {Waals} heterostructures},\ }\href {https://doi.org/10.1103/PhysRevB.103.L041406} {\bibfield  {journal} {\bibinfo  {journal} {Physical Review B}\ }\textbf {\bibinfo {volume} {103}},\ \bibinfo {pages} {L041406} (\bibinfo {year} {2021})}\BibitemShut {NoStop}%
\bibitem [{\citenamefont {Nielsen}\ \emph {et~al.}(2023)\citenamefont {Nielsen}, \citenamefont {da~Cruz}, \citenamefont {Torche},\ and\ \citenamefont {Bester}}]{nielsen_accurate_2023}%
  \BibitemOpen
  \bibfield  {author} {\bibinfo {author} {\bibfnamefont {C.~E.~M.}\ \bibnamefont {Nielsen}}, \bibinfo {author} {\bibfnamefont {M.}~\bibnamefont {da~Cruz}}, \bibinfo {author} {\bibfnamefont {A.}~\bibnamefont {Torche}},\ and\ \bibinfo {author} {\bibfnamefont {G.}~\bibnamefont {Bester}},\ }\bibfield  {title} {\bibinfo {title} {Accurate force-field methodology capturing atomic reconstructions in transition metal dichalcogenide moiré system},\ }\href {https://doi.org/10.1103/PhysRevB.108.045402} {\bibfield  {journal} {\bibinfo  {journal} {Physical Review B}\ }\textbf {\bibinfo {volume} {108}},\ \bibinfo {pages} {045402} (\bibinfo {year} {2023})}\BibitemShut {NoStop}%
\bibitem [{\citenamefont {Seyler}\ \emph {et~al.}(2019)\citenamefont {Seyler}, \citenamefont {Rivera}, \citenamefont {Yu}, \citenamefont {Wilson}, \citenamefont {Ray}, \citenamefont {Mandrus}, \citenamefont {Yan}, \citenamefont {Yao},\ and\ \citenamefont {Xu}}]{seyler_signatures_2019}%
  \BibitemOpen
  \bibfield  {author} {\bibinfo {author} {\bibfnamefont {K.~L.}\ \bibnamefont {Seyler}}, \bibinfo {author} {\bibfnamefont {P.}~\bibnamefont {Rivera}}, \bibinfo {author} {\bibfnamefont {H.}~\bibnamefont {Yu}}, \bibinfo {author} {\bibfnamefont {N.~P.}\ \bibnamefont {Wilson}}, \bibinfo {author} {\bibfnamefont {E.~L.}\ \bibnamefont {Ray}}, \bibinfo {author} {\bibfnamefont {D.~G.}\ \bibnamefont {Mandrus}}, \bibinfo {author} {\bibfnamefont {J.}~\bibnamefont {Yan}}, \bibinfo {author} {\bibfnamefont {W.}~\bibnamefont {Yao}},\ and\ \bibinfo {author} {\bibfnamefont {X.}~\bibnamefont {Xu}},\ }\bibfield  {title} {\bibinfo {title} {Signatures of moiré-trapped valley excitons in {MoSe}$_2$/{WSe}$_2$ heterobilayers},\ }\href {https://doi.org/10.1038/s41586-019-0957-1} {\bibfield  {journal} {\bibinfo  {journal} {Nature}\ }\textbf {\bibinfo {volume} {567}},\ \bibinfo {pages} {66} (\bibinfo {year} {2019})}\BibitemShut {NoStop}%
\bibitem [{\citenamefont {Zhang}\ \emph {et~al.}(2020)\citenamefont {Zhang}, \citenamefont {Zhang}, \citenamefont {Wu}, \citenamefont {Wang}, \citenamefont {Gogna}, \citenamefont {Hou}, \citenamefont {Watanabe}, \citenamefont {Taniguchi}, \citenamefont {Kulkarni}, \citenamefont {Kuo}, \citenamefont {Forrest},\ and\ \citenamefont {Deng}}]{zhang_twist-angle_2020}%
  \BibitemOpen
  \bibfield  {author} {\bibinfo {author} {\bibfnamefont {L.}~\bibnamefont {Zhang}}, \bibinfo {author} {\bibfnamefont {Z.}~\bibnamefont {Zhang}}, \bibinfo {author} {\bibfnamefont {F.}~\bibnamefont {Wu}}, \bibinfo {author} {\bibfnamefont {D.}~\bibnamefont {Wang}}, \bibinfo {author} {\bibfnamefont {R.}~\bibnamefont {Gogna}}, \bibinfo {author} {\bibfnamefont {S.}~\bibnamefont {Hou}}, \bibinfo {author} {\bibfnamefont {K.}~\bibnamefont {Watanabe}}, \bibinfo {author} {\bibfnamefont {T.}~\bibnamefont {Taniguchi}}, \bibinfo {author} {\bibfnamefont {K.}~\bibnamefont {Kulkarni}}, \bibinfo {author} {\bibfnamefont {T.}~\bibnamefont {Kuo}}, \bibinfo {author} {\bibfnamefont {S.~R.}\ \bibnamefont {Forrest}},\ and\ \bibinfo {author} {\bibfnamefont {H.}~\bibnamefont {Deng}},\ }\bibfield  {title} {\bibinfo {title} {Twist-angle dependence of moiré excitons in {WS}$_2$/{MoSe}$_2$ heterobilayers},\ }\href {https://doi.org/10.1038/s41467-020-19466-6} {\bibfield  {journal} {\bibinfo  {journal} {Nature Communications}\ }\textbf
  {\bibinfo {volume} {11}},\ \bibinfo {pages} {5888} (\bibinfo {year} {2020})}\BibitemShut {NoStop}%
\bibitem [{\citenamefont {Sigl}\ \emph {et~al.}(2020)\citenamefont {Sigl}, \citenamefont {Sigger}, \citenamefont {Kronowetter}, \citenamefont {Kiemle}, \citenamefont {Klein}, \citenamefont {Watanabe}, \citenamefont {Taniguchi}, \citenamefont {Finley}, \citenamefont {Wurstbauer},\ and\ \citenamefont {Holleitner}}]{sigl_signatures_2020}%
  \BibitemOpen
  \bibfield  {author} {\bibinfo {author} {\bibfnamefont {L.}~\bibnamefont {Sigl}}, \bibinfo {author} {\bibfnamefont {F.}~\bibnamefont {Sigger}}, \bibinfo {author} {\bibfnamefont {F.}~\bibnamefont {Kronowetter}}, \bibinfo {author} {\bibfnamefont {J.}~\bibnamefont {Kiemle}}, \bibinfo {author} {\bibfnamefont {J.}~\bibnamefont {Klein}}, \bibinfo {author} {\bibfnamefont {K.}~\bibnamefont {Watanabe}}, \bibinfo {author} {\bibfnamefont {T.}~\bibnamefont {Taniguchi}}, \bibinfo {author} {\bibfnamefont {J.~J.}\ \bibnamefont {Finley}}, \bibinfo {author} {\bibfnamefont {U.}~\bibnamefont {Wurstbauer}},\ and\ \bibinfo {author} {\bibfnamefont {A.~W.}\ \bibnamefont {Holleitner}},\ }\bibfield  {title} {\bibinfo {title} {Signatures of a degenerate many-body state of interlayer excitons in a van der {Waals} heterostack},\ }\href {https://doi.org/10.1103/PhysRevResearch.2.042044} {\bibfield  {journal} {\bibinfo  {journal} {Physical Review Research}\ }\textbf {\bibinfo {volume} {2}},\ \bibinfo {pages} {042044} (\bibinfo {year}
  {2020})}\BibitemShut {NoStop}%
\bibitem [{\citenamefont {Troue}\ \emph {et~al.}(2023)\citenamefont {Troue}, \citenamefont {Figueiredo}, \citenamefont {Sigl}, \citenamefont {Paspalides}, \citenamefont {Katzer}, \citenamefont {Taniguchi}, \citenamefont {Watanabe}, \citenamefont {Selig}, \citenamefont {Knorr}, \citenamefont {Wurstbauer},\ and\ \citenamefont {Holleitner}}]{troue_extended_2023}%
  \BibitemOpen
  \bibfield  {author} {\bibinfo {author} {\bibfnamefont {M.}~\bibnamefont {Troue}}, \bibinfo {author} {\bibfnamefont {J.}~\bibnamefont {Figueiredo}}, \bibinfo {author} {\bibfnamefont {L.}~\bibnamefont {Sigl}}, \bibinfo {author} {\bibfnamefont {C.}~\bibnamefont {Paspalides}}, \bibinfo {author} {\bibfnamefont {M.}~\bibnamefont {Katzer}}, \bibinfo {author} {\bibfnamefont {T.}~\bibnamefont {Taniguchi}}, \bibinfo {author} {\bibfnamefont {K.}~\bibnamefont {Watanabe}}, \bibinfo {author} {\bibfnamefont {M.}~\bibnamefont {Selig}}, \bibinfo {author} {\bibfnamefont {A.}~\bibnamefont {Knorr}}, \bibinfo {author} {\bibfnamefont {U.}~\bibnamefont {Wurstbauer}},\ and\ \bibinfo {author} {\bibfnamefont {A.~W.}\ \bibnamefont {Holleitner}},\ }\bibfield  {title} {\bibinfo {title} {Extended {Spatial} {Coherence} of {Interlayer} {Excitons} in {MoSe}$_2$/{WSe}$_2$ {Heterobilayers}},\ }\href {https://doi.org/10.1103/PhysRevLett.131.036902} {\bibfield  {journal} {\bibinfo  {journal} {Physical Review Letters}\ }\textbf {\bibinfo
  {volume} {131}},\ \bibinfo {pages} {036902} (\bibinfo {year} {2023})}\BibitemShut {NoStop}%
\bibitem [{\citenamefont {Richter}(2024)}]{richter_theory_2024}%
  \BibitemOpen
  \bibfield  {author} {\bibinfo {author} {\bibfnamefont {M.}~\bibnamefont {Richter}},\ }\bibfield  {title} {\bibinfo {title} {Theory of interlayer exciton dynamics in two-dimensional transition metal dichalcogenide heterolayers under the influence of strain reconstruction and disorder},\ }\href {https://doi.org/10.1103/PhysRevB.109.125308} {\bibfield  {journal} {\bibinfo  {journal} {Physical Review B}\ }\textbf {\bibinfo {volume} {109}},\ \bibinfo {pages} {125308} (\bibinfo {year} {2024})}\BibitemShut {NoStop}%
\bibitem [{\citenamefont {Enaldiev}\ \emph {et~al.}(2020)\citenamefont {Enaldiev}, \citenamefont {Zólyomi}, \citenamefont {Yelgel}, \citenamefont {Magorrian},\ and\ \citenamefont {Fal’ko}}]{enaldiev_stacking_2020}%
  \BibitemOpen
  \bibfield  {author} {\bibinfo {author} {\bibfnamefont {V.}~\bibnamefont {Enaldiev}}, \bibinfo {author} {\bibfnamefont {V.}~\bibnamefont {Zólyomi}}, \bibinfo {author} {\bibfnamefont {C.}~\bibnamefont {Yelgel}}, \bibinfo {author} {\bibfnamefont {S.}~\bibnamefont {Magorrian}},\ and\ \bibinfo {author} {\bibfnamefont {V.}~\bibnamefont {Fal’ko}},\ }\bibfield  {title} {\bibinfo {title} {Stacking {Domains} and {Dislocation} {Networks} in {Marginally} {Twisted} {Bilayers} of {Transition} {Metal} {Dichalcogenides}},\ }\href {https://doi.org/10.1103/PhysRevLett.124.206101} {\bibfield  {journal} {\bibinfo  {journal} {Physical Review Letters}\ }\textbf {\bibinfo {volume} {124}},\ \bibinfo {pages} {206101} (\bibinfo {year} {2020})}\BibitemShut {NoStop}%
\bibitem [{\citenamefont {Singh}\ \emph {et~al.}(2017)\citenamefont {Singh}, \citenamefont {Richter}, \citenamefont {Moody}, \citenamefont {Siemens}, \citenamefont {Li},\ and\ \citenamefont {Cundiff}}]{singh_localization_2017}%
  \BibitemOpen
  \bibfield  {author} {\bibinfo {author} {\bibfnamefont {R.}~\bibnamefont {Singh}}, \bibinfo {author} {\bibfnamefont {M.}~\bibnamefont {Richter}}, \bibinfo {author} {\bibfnamefont {G.}~\bibnamefont {Moody}}, \bibinfo {author} {\bibfnamefont {M.~E.}\ \bibnamefont {Siemens}}, \bibinfo {author} {\bibfnamefont {H.}~\bibnamefont {Li}},\ and\ \bibinfo {author} {\bibfnamefont {S.~T.}\ \bibnamefont {Cundiff}},\ }\bibfield  {title} {\bibinfo {title} {Localization dynamics of excitons in disordered semiconductor quantum wells},\ }\href {https://doi.org/10.1103/PhysRevB.95.235307} {\bibfield  {journal} {\bibinfo  {journal} {Physical Review B}\ }\textbf {\bibinfo {volume} {95}},\ \bibinfo {pages} {235307} (\bibinfo {year} {2017})}\BibitemShut {NoStop}%
\bibitem [{\citenamefont {Richter}\ \emph {et~al.}(2018)\citenamefont {Richter}, \citenamefont {Singh}, \citenamefont {Siemens},\ and\ \citenamefont {Cundiff}}]{richter_deconvolution_2018}%
  \BibitemOpen
  \bibfield  {author} {\bibinfo {author} {\bibfnamefont {M.}~\bibnamefont {Richter}}, \bibinfo {author} {\bibfnamefont {R.}~\bibnamefont {Singh}}, \bibinfo {author} {\bibfnamefont {M.}~\bibnamefont {Siemens}},\ and\ \bibinfo {author} {\bibfnamefont {S.~T.}\ \bibnamefont {Cundiff}},\ }\bibfield  {title} {\bibinfo {title} {Deconvolution of optical multidimensional coherent spectra},\ }\href {https://doi.org/10.1126/sciadv.aar7697} {\bibfield  {journal} {\bibinfo  {journal} {Science Advances}\ }\textbf {\bibinfo {volume} {4}},\ \bibinfo {pages} {eaar7697} (\bibinfo {year} {2018})}\BibitemShut {NoStop}%
\bibitem [{\citenamefont {Khatibi}\ \emph {et~al.}(2018)\citenamefont {Khatibi}, \citenamefont {Feierabend}, \citenamefont {Selig}, \citenamefont {Brem}, \citenamefont {Linderälv}, \citenamefont {Erhart},\ and\ \citenamefont {Malic}}]{khatibi_impact_2018}%
  \BibitemOpen
  \bibfield  {author} {\bibinfo {author} {\bibfnamefont {Z.}~\bibnamefont {Khatibi}}, \bibinfo {author} {\bibfnamefont {M.}~\bibnamefont {Feierabend}}, \bibinfo {author} {\bibfnamefont {M.}~\bibnamefont {Selig}}, \bibinfo {author} {\bibfnamefont {S.}~\bibnamefont {Brem}}, \bibinfo {author} {\bibfnamefont {C.}~\bibnamefont {Linderälv}}, \bibinfo {author} {\bibfnamefont {P.}~\bibnamefont {Erhart}},\ and\ \bibinfo {author} {\bibfnamefont {E.}~\bibnamefont {Malic}},\ }\bibfield  {title} {\bibinfo {title} {Impact of strain on the excitonic linewidth in transition metal dichalcogenides},\ }\href {https://doi.org/10.1088/2053-1583/aae953} {\bibfield  {journal} {\bibinfo  {journal} {2D Materials}\ }\textbf {\bibinfo {volume} {6}},\ \bibinfo {pages} {015015} (\bibinfo {year} {2018})}\BibitemShut {NoStop}%
\bibitem [{\citenamefont {Wu}\ \emph {et~al.}(2018)\citenamefont {Wu}, \citenamefont {Lovorn},\ and\ \citenamefont {MacDonald}}]{wu_theory_2018}%
  \BibitemOpen
  \bibfield  {author} {\bibinfo {author} {\bibfnamefont {F.}~\bibnamefont {Wu}}, \bibinfo {author} {\bibfnamefont {T.}~\bibnamefont {Lovorn}},\ and\ \bibinfo {author} {\bibfnamefont {A.~H.}\ \bibnamefont {MacDonald}},\ }\bibfield  {title} {\bibinfo {title} {Theory of optical absorption by interlayer excitons in transition metal dichalcogenide heterobilayers},\ }\href {https://doi.org/10.1103/PhysRevB.97.035306} {\bibfield  {journal} {\bibinfo  {journal} {Physical Review B}\ }\textbf {\bibinfo {volume} {97}},\ \bibinfo {pages} {035306} (\bibinfo {year} {2018})}\BibitemShut {NoStop}%
\bibitem [{\citenamefont {Tran}\ \emph {et~al.}(2019)\citenamefont {Tran}, \citenamefont {Moody}, \citenamefont {Wu}, \citenamefont {Lu}, \citenamefont {Choi}, \citenamefont {Kim}, \citenamefont {Rai}, \citenamefont {Sanchez}, \citenamefont {Quan}, \citenamefont {Singh}, \citenamefont {Embley}, \citenamefont {Zepeda}, \citenamefont {Campbell}, \citenamefont {Autry}, \citenamefont {Taniguchi}, \citenamefont {Watanabe}, \citenamefont {Lu}, \citenamefont {Banerjee}, \citenamefont {Silverman}, \citenamefont {Kim}, \citenamefont {Tutuc}, \citenamefont {Yang}, \citenamefont {MacDonald},\ and\ \citenamefont {Li}}]{tran_evidence_2019}%
  \BibitemOpen
  \bibfield  {author} {\bibinfo {author} {\bibfnamefont {K.}~\bibnamefont {Tran}}, \bibinfo {author} {\bibfnamefont {G.}~\bibnamefont {Moody}}, \bibinfo {author} {\bibfnamefont {F.}~\bibnamefont {Wu}}, \bibinfo {author} {\bibfnamefont {X.}~\bibnamefont {Lu}}, \bibinfo {author} {\bibfnamefont {J.}~\bibnamefont {Choi}}, \bibinfo {author} {\bibfnamefont {K.}~\bibnamefont {Kim}}, \bibinfo {author} {\bibfnamefont {A.}~\bibnamefont {Rai}}, \bibinfo {author} {\bibfnamefont {D.~A.}\ \bibnamefont {Sanchez}}, \bibinfo {author} {\bibfnamefont {J.}~\bibnamefont {Quan}}, \bibinfo {author} {\bibfnamefont {A.}~\bibnamefont {Singh}}, \bibinfo {author} {\bibfnamefont {J.}~\bibnamefont {Embley}}, \bibinfo {author} {\bibfnamefont {A.}~\bibnamefont {Zepeda}}, \bibinfo {author} {\bibfnamefont {M.}~\bibnamefont {Campbell}}, \bibinfo {author} {\bibfnamefont {T.}~\bibnamefont {Autry}}, \bibinfo {author} {\bibfnamefont {T.}~\bibnamefont {Taniguchi}}, \bibinfo {author} {\bibfnamefont {K.}~\bibnamefont {Watanabe}}, \bibinfo {author}
  {\bibfnamefont {N.}~\bibnamefont {Lu}}, \bibinfo {author} {\bibfnamefont {S.~K.}\ \bibnamefont {Banerjee}}, \bibinfo {author} {\bibfnamefont {K.~L.}\ \bibnamefont {Silverman}}, \bibinfo {author} {\bibfnamefont {S.}~\bibnamefont {Kim}}, \bibinfo {author} {\bibfnamefont {E.}~\bibnamefont {Tutuc}}, \bibinfo {author} {\bibfnamefont {L.}~\bibnamefont {Yang}}, \bibinfo {author} {\bibfnamefont {A.~H.}\ \bibnamefont {MacDonald}},\ and\ \bibinfo {author} {\bibfnamefont {X.}~\bibnamefont {Li}},\ }\bibfield  {title} {\bibinfo {title} {Evidence for moiré excitons in van der {Waals} heterostructures},\ }\href {https://doi.org/10.1038/s41586-019-0975-z} {\bibfield  {journal} {\bibinfo  {journal} {Nature}\ }\textbf {\bibinfo {volume} {567}},\ \bibinfo {pages} {71} (\bibinfo {year} {2019})}\BibitemShut {NoStop}%
\bibitem [{\citenamefont {Ruiz-Tijerina}\ \emph {et~al.}(2020)\citenamefont {Ruiz-Tijerina}, \citenamefont {Soltero},\ and\ \citenamefont {Mireles}}]{ruiz-tijerina_theory_2020}%
  \BibitemOpen
  \bibfield  {author} {\bibinfo {author} {\bibfnamefont {D.~A.}\ \bibnamefont {Ruiz-Tijerina}}, \bibinfo {author} {\bibfnamefont {I.}~\bibnamefont {Soltero}},\ and\ \bibinfo {author} {\bibfnamefont {F.}~\bibnamefont {Mireles}},\ }\bibfield  {title} {\bibinfo {title} {Theory of moiré localized excitons in transition metal dichalcogenide heterobilayers},\ }\href {https://doi.org/10.1103/PhysRevB.102.195403} {\bibfield  {journal} {\bibinfo  {journal} {Physical Review B}\ }\textbf {\bibinfo {volume} {102}},\ \bibinfo {pages} {195403} (\bibinfo {year} {2020})}\BibitemShut {NoStop}%
\bibitem [{\citenamefont {Fallahazad}\ \emph {et~al.}(2016)\citenamefont {Fallahazad}, \citenamefont {Movva}, \citenamefont {Kim}, \citenamefont {Larentis}, \citenamefont {Taniguchi}, \citenamefont {Watanabe}, \citenamefont {Banerjee},\ and\ \citenamefont {Tutuc}}]{fallahazad_shubnikov--haas_2016}%
  \BibitemOpen
  \bibfield  {author} {\bibinfo {author} {\bibfnamefont {B.}~\bibnamefont {Fallahazad}}, \bibinfo {author} {\bibfnamefont {H.~C.}\ \bibnamefont {Movva}}, \bibinfo {author} {\bibfnamefont {K.}~\bibnamefont {Kim}}, \bibinfo {author} {\bibfnamefont {S.}~\bibnamefont {Larentis}}, \bibinfo {author} {\bibfnamefont {T.}~\bibnamefont {Taniguchi}}, \bibinfo {author} {\bibfnamefont {K.}~\bibnamefont {Watanabe}}, \bibinfo {author} {\bibfnamefont {S.~K.}\ \bibnamefont {Banerjee}},\ and\ \bibinfo {author} {\bibfnamefont {E.}~\bibnamefont {Tutuc}},\ }\bibfield  {title} {\bibinfo {title} {Shubnikov--de {Haas} {Oscillations} of {High}-{Mobility} {Holes} in {Monolayer} and {Bilayer} {WSe}$_2$: {Landau} {Level} {Degeneracy}, {Effective} {Mass}, and {Negative} {Compressibility}},\ }\href {https://doi.org/10.1103/PhysRevLett.116.086601} {\bibfield  {journal} {\bibinfo  {journal} {Physical Review Letters}\ }\textbf {\bibinfo {volume} {116}},\ \bibinfo {pages} {086601} (\bibinfo {year} {2016})}\BibitemShut {NoStop}%
\bibitem [{\citenamefont {Mostaani}\ \emph {et~al.}(2017)\citenamefont {Mostaani}, \citenamefont {Szyniszewski}, \citenamefont {Price}, \citenamefont {Maezono}, \citenamefont {Danovich}, \citenamefont {Hunt}, \citenamefont {Drummond},\ and\ \citenamefont {Fal'ko}}]{mostaani_diffusion_2017}%
  \BibitemOpen
  \bibfield  {author} {\bibinfo {author} {\bibfnamefont {E.}~\bibnamefont {Mostaani}}, \bibinfo {author} {\bibfnamefont {M.}~\bibnamefont {Szyniszewski}}, \bibinfo {author} {\bibfnamefont {C.~H.}\ \bibnamefont {Price}}, \bibinfo {author} {\bibfnamefont {R.}~\bibnamefont {Maezono}}, \bibinfo {author} {\bibfnamefont {M.}~\bibnamefont {Danovich}}, \bibinfo {author} {\bibfnamefont {R.~J.}\ \bibnamefont {Hunt}}, \bibinfo {author} {\bibfnamefont {N.~D.}\ \bibnamefont {Drummond}},\ and\ \bibinfo {author} {\bibfnamefont {V.~I.}\ \bibnamefont {Fal'ko}},\ }\bibfield  {title} {\bibinfo {title} {Diffusion quantum {Monte} {Carlo} study of excitonic complexes in two-dimensional transition-metal dichalcogenides},\ }\href {https://doi.org/10.1103/PhysRevB.96.075431} {\bibfield  {journal} {\bibinfo  {journal} {Physical Review B}\ }\textbf {\bibinfo {volume} {96}},\ \bibinfo {pages} {075431} (\bibinfo {year} {2017})}\BibitemShut {NoStop}%
\bibitem [{\citenamefont {Jin}\ \emph {et~al.}(2014)\citenamefont {Jin}, \citenamefont {Li}, \citenamefont {Mullen},\ and\ \citenamefont {Kim}}]{jin_intrinsic_2014}%
  \BibitemOpen
  \bibfield  {author} {\bibinfo {author} {\bibfnamefont {Z.}~\bibnamefont {Jin}}, \bibinfo {author} {\bibfnamefont {X.}~\bibnamefont {Li}}, \bibinfo {author} {\bibfnamefont {J.~T.}\ \bibnamefont {Mullen}},\ and\ \bibinfo {author} {\bibfnamefont {K.~W.}\ \bibnamefont {Kim}},\ }\bibfield  {title} {\bibinfo {title} {Intrinsic transport properties of electrons and holes in monolayer transition-metal dichalcogenides},\ }\href {https://doi.org/10.1103/PhysRevB.90.045422} {\bibfield  {journal} {\bibinfo  {journal} {Physical Review B}\ }\textbf {\bibinfo {volume} {90}},\ \bibinfo {pages} {045422} (\bibinfo {year} {2014})}\BibitemShut {NoStop}%
\bibitem [{\citenamefont {Shree}\ \emph {et~al.}(2018)\citenamefont {Shree}, \citenamefont {Semina}, \citenamefont {Robert}, \citenamefont {Han}, \citenamefont {Amand}, \citenamefont {Balocchi}, \citenamefont {Manca}, \citenamefont {Courtade}, \citenamefont {Marie}, \citenamefont {Taniguchi}, \citenamefont {Watanabe}, \citenamefont {Glazov},\ and\ \citenamefont {Urbaszek}}]{shree_observation_2018}%
  \BibitemOpen
  \bibfield  {author} {\bibinfo {author} {\bibfnamefont {S.}~\bibnamefont {Shree}}, \bibinfo {author} {\bibfnamefont {M.}~\bibnamefont {Semina}}, \bibinfo {author} {\bibfnamefont {C.}~\bibnamefont {Robert}}, \bibinfo {author} {\bibfnamefont {B.}~\bibnamefont {Han}}, \bibinfo {author} {\bibfnamefont {T.}~\bibnamefont {Amand}}, \bibinfo {author} {\bibfnamefont {A.}~\bibnamefont {Balocchi}}, \bibinfo {author} {\bibfnamefont {M.}~\bibnamefont {Manca}}, \bibinfo {author} {\bibfnamefont {E.}~\bibnamefont {Courtade}}, \bibinfo {author} {\bibfnamefont {X.}~\bibnamefont {Marie}}, \bibinfo {author} {\bibfnamefont {T.}~\bibnamefont {Taniguchi}}, \bibinfo {author} {\bibfnamefont {K.}~\bibnamefont {Watanabe}}, \bibinfo {author} {\bibfnamefont {M.~M.}\ \bibnamefont {Glazov}},\ and\ \bibinfo {author} {\bibfnamefont {B.}~\bibnamefont {Urbaszek}},\ }\bibfield  {title} {\bibinfo {title} {Observation of exciton-phonon coupling in {MoSe}$_2$ monolayers},\ }\href {https://doi.org/10.1103/PhysRevB.98.035302} {\bibfield  {journal}
  {\bibinfo  {journal} {Physical Review B}\ }\textbf {\bibinfo {volume} {98}},\ \bibinfo {pages} {035302} (\bibinfo {year} {2018})}\BibitemShut {NoStop}%
\bibitem [{\citenamefont {Zimmermann}\ \emph {et~al.}(2003)\citenamefont {Zimmermann}, \citenamefont {Runge},\ and\ \citenamefont {Savona}}]{zimmermann_chapter_2003}%
  \BibitemOpen
  \bibfield  {author} {\bibinfo {author} {\bibfnamefont {R.}~\bibnamefont {Zimmermann}}, \bibinfo {author} {\bibfnamefont {E.}~\bibnamefont {Runge}},\ and\ \bibinfo {author} {\bibfnamefont {V.}~\bibnamefont {Savona}},\ }\bibfield  {title} {\bibinfo {title} {Chapter 4 - {Theory} of resonant secondary emission: {Rayleigh} scattering versus luminescence},\ }in\ \href {https://doi.org/10.1016/B978-012682225-0/50005-5} {\emph {\bibinfo {booktitle} {Quantum {Coherence} {Correlation} and {Decoherence} in {Semiconductor} {Nanostructures}}}},\ \bibinfo {editor} {edited by\ \bibinfo {editor} {\bibfnamefont {T.}~\bibnamefont {Takagahara}}}\ (\bibinfo  {publisher} {Academic Press},\ \bibinfo {address} {San Diego},\ \bibinfo {year} {2003})\ pp.\ \bibinfo {pages} {89--165}\BibitemShut {NoStop}%
\bibitem [{\citenamefont {Hermann}(2012)}]{hermann_periodic_2012}%
  \BibitemOpen
  \bibfield  {author} {\bibinfo {author} {\bibfnamefont {K.}~\bibnamefont {Hermann}},\ }\bibfield  {title} {\bibinfo {title} {Periodic overlayers and moiré patterns: theoretical studies of geometric properties},\ }\href {https://doi.org/10.1088/0953-8984/24/31/314210} {\bibfield  {journal} {\bibinfo  {journal} {Journal of Physics: Condensed Matter}\ }\textbf {\bibinfo {volume} {24}},\ \bibinfo {pages} {314210} (\bibinfo {year} {2012})}\BibitemShut {NoStop}%
\bibitem [{\citenamefont {Urbach}(1953)}]{urbach_long-wavelength_1953}%
  \BibitemOpen
  \bibfield  {author} {\bibinfo {author} {\bibfnamefont {F.}~\bibnamefont {Urbach}},\ }\bibfield  {title} {\bibinfo {title} {The {Long}-{Wavelength} {Edge} of {Photographic} {Sensitivity} and of the {Electronic} {Absorption} of {Solids}},\ }\href {https://doi.org/10.1103/PhysRev.92.1324} {\bibfield  {journal} {\bibinfo  {journal} {Physical Review}\ }\textbf {\bibinfo {volume} {92}},\ \bibinfo {pages} {1324} (\bibinfo {year} {1953})}\BibitemShut {NoStop}%
\bibitem [{\citenamefont {Piccardo}\ \emph {et~al.}(2017)\citenamefont {Piccardo}, \citenamefont {Li}, \citenamefont {Wu}, \citenamefont {Speck}, \citenamefont {Bonef}, \citenamefont {Farrell}, \citenamefont {Filoche}, \citenamefont {Martinelli}, \citenamefont {Peretti},\ and\ \citenamefont {Weisbuch}}]{piccardo_localization_2017}%
  \BibitemOpen
  \bibfield  {author} {\bibinfo {author} {\bibfnamefont {M.}~\bibnamefont {Piccardo}}, \bibinfo {author} {\bibfnamefont {C.-K.}\ \bibnamefont {Li}}, \bibinfo {author} {\bibfnamefont {Y.-R.}\ \bibnamefont {Wu}}, \bibinfo {author} {\bibfnamefont {J.~S.}\ \bibnamefont {Speck}}, \bibinfo {author} {\bibfnamefont {B.}~\bibnamefont {Bonef}}, \bibinfo {author} {\bibfnamefont {R.~M.}\ \bibnamefont {Farrell}}, \bibinfo {author} {\bibfnamefont {M.}~\bibnamefont {Filoche}}, \bibinfo {author} {\bibfnamefont {L.}~\bibnamefont {Martinelli}}, \bibinfo {author} {\bibfnamefont {J.}~\bibnamefont {Peretti}},\ and\ \bibinfo {author} {\bibfnamefont {C.}~\bibnamefont {Weisbuch}},\ }\bibfield  {title} {\bibinfo {title} {Localization landscape theory of disorder in semiconductors. {II}. {Urbach} tails of disordered quantum well layers},\ }\href {https://doi.org/10.1103/PhysRevB.95.144205} {\bibfield  {journal} {\bibinfo  {journal} {Physical Review B}\ }\textbf {\bibinfo {volume} {95}},\ \bibinfo {pages} {144205} (\bibinfo {year}
  {2017})}\BibitemShut {NoStop}%
\bibitem [{\citenamefont {Katzer}\ \emph {et~al.}(2023)\citenamefont {Katzer}, \citenamefont {Selig}, \citenamefont {Sigl}, \citenamefont {Troue}, \citenamefont {Figueiredo}, \citenamefont {Kiemle}, \citenamefont {Sigger}, \citenamefont {Wurstbauer}, \citenamefont {Holleitner},\ and\ \citenamefont {Knorr}}]{katzer_exciton-phonon_2023}%
  \BibitemOpen
  \bibfield  {author} {\bibinfo {author} {\bibfnamefont {M.}~\bibnamefont {Katzer}}, \bibinfo {author} {\bibfnamefont {M.}~\bibnamefont {Selig}}, \bibinfo {author} {\bibfnamefont {L.}~\bibnamefont {Sigl}}, \bibinfo {author} {\bibfnamefont {M.}~\bibnamefont {Troue}}, \bibinfo {author} {\bibfnamefont {J.}~\bibnamefont {Figueiredo}}, \bibinfo {author} {\bibfnamefont {J.}~\bibnamefont {Kiemle}}, \bibinfo {author} {\bibfnamefont {F.}~\bibnamefont {Sigger}}, \bibinfo {author} {\bibfnamefont {U.}~\bibnamefont {Wurstbauer}}, \bibinfo {author} {\bibfnamefont {A.~W.}\ \bibnamefont {Holleitner}},\ and\ \bibinfo {author} {\bibfnamefont {A.}~\bibnamefont {Knorr}},\ }\bibfield  {title} {\bibinfo {title} {Exciton-phonon scattering: {Competition} between the bosonic and fermionic nature of bound electron-hole pairs},\ }\href {https://doi.org/10.1103/PhysRevB.108.L121102} {\bibfield  {journal} {\bibinfo  {journal} {Physical Review B}\ }\textbf {\bibinfo {volume} {108}},\ \bibinfo {pages} {L121102} (\bibinfo {year}
  {2023})}\BibitemShut {NoStop}%
\end{thebibliography}%

\clearpage
\includepdf[pages=1]{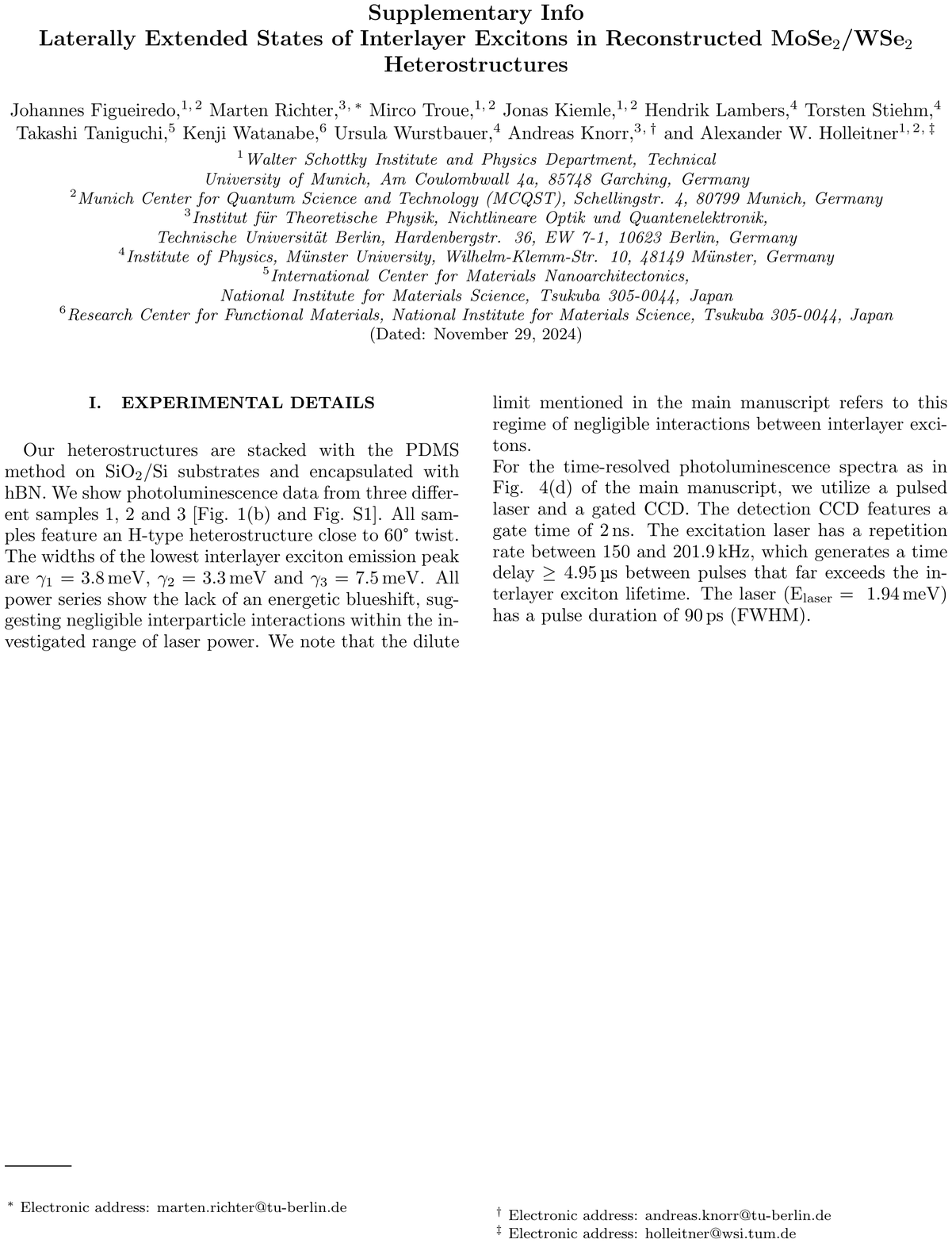}
\clearpage
\includepdf[pages=2]{SI}
\clearpage
\includepdf[pages=3]{SI}
\clearpage
\includepdf[pages=4]{SI}

\end{document}